\documentclass[mathleft]{an}
\usepackage{graphicx}
\usepackage{times}
\begin{document}

\Pagespan{789}{}
\Yearpublication{2006}%
\Yearsubmission{2005}%
\Month{11}%
\Volume{999}%
\Issue{88}%

\title{Search for a circum-planetary material and orbital period variations of short-period Kepler exoplanet candidates}

\author{Z. Garai\inst{1}\fnmsep\thanks{Corresponding author:
  \email{zgarai@ta3.sk}\newline}
\and G. Zhou\inst{3}
\and J. Budaj\inst{3}\fnmsep\inst{1}
\and R.F. Stellingwerf\inst{2}
}
\titlerunning{Search for a circum-planetary material and orbital period
variations\dots}
\authorrunning{Z. Garai et al.}
\institute{
Astronomical Institute, Slovak Academy of Sciences, 059 60 Tatranska
Lomnica, Slovak Republic
\and 
Stellingwerf Consulting, 11033 Mathis Mtn Rd SE, Huntsville, AL 35803, United States of America
\and
Research School of Astronomy and Astrophysics, Australian National
University, Canberra, ACT 2611, Australia}
\received{dd mmm yyyy}
\accepted{dd mmm yyyy}
\publonline{dd mmm yyyy}


\abstract{%
  \\A unique short-period ($P = 0.65356(1)$ days) Mercury-size Kepler exoplanet candidate KIC012557548b has been discovered
  recently by \textit{Rappaport et al. (2012)}. This object is 
  a transiting disintegrating exoplanet with a circum-planetary material -- comet-like tail.
  Close-in exoplanets, like KIC012557548b, are subjected to the greatest
  planet-star interactions. This interaction may have various forms.
  In certain cases it may cause formation of the comet-like tail.
  Strong interaction with the host star, 
  and/or presence of an additional planet may lead to variations in 
  the orbital period of the planet.
  \\Our main aim is to search for comet-like tails similar to KIC012557548b
  and for long-term orbital period variations. We are curious about frequency of comet-like tail 
  formation among short-period Kepler exoplanet candidates.
  We concentrate on a sample of 20 close-in candidates with a period similar to KIC012557548b from the Kepler
  mission. 
  \\We first improved the preliminary orbital periods and obtained the transit light curves.
  Subsequently we searched for the signatures of a circum-planetary material
  in these light curves. For this purpose the final transit light curve of each planet was 
  fitted with a theoretical light curve, and the residuals were examined
  for abnormalities. We then searched for possible long-term 
  changes of the orbital periods using the method of phase dispersion 
  minimization. 
  \\In 8 cases out of 20 we found some interesting peculiarities, but none 
  of the exoplanet candidates showed signs of a comet-like tail. It seems that the frequency of comet-like tail formation among short-period
  Kepler exoplanet candidates is very low. We searched for comet-like tails based on the period criterion. 
  Based on our results we can conclude that the short-period criterion is not enough to cause comet-like tail formation.
  This result is in agreement with the theory of the thermal wind and planet evaporation (Perez-Becker \& Chiang 2013). 
  We also found 3 cases of candidates which showed some changes of the orbital period. Based on our results we can see that orbital period changes are
  not caused by comet-like tail disintegration processes, but rather by possible massive outer companions.}

\maketitle


\section{Introduction}
Close-in exoplanets are subjected to the greatest planet-star interactions.
It may have various forms.
(1) Heavy irradiation changes the atmospheric structure and 
creates a deep temperature plateau or a stratosphere
(Hubeny, Burrows \& Sudarsky 2003; Burrows, Budaj \& Hubeny 2008; 
Knutson et al. 2008; Fortney et al. 2008).
(2) Strong irradiation drives the mass loss from the planet
(Burrows \& Lunine 1995; Guillot et al. 1996).
It was detected in HD 209458b (Vidal-Madjar et al. 2003, 2004), 
and in HD 189733b (Lecavelier des Etangs et al. 2010; 
Bourrier et al. 2013). 
Several theoretical studies were devoted to this subject
(e.g. Yelle 2004; Tian et al. 2005; Hubbard et al. 2007;
Lopez, Fortney \& Miller 2012).
\textit{Kurokawa \& Kaltenegger (2013)} developed a combined model 
of atmospheric mass loss calculation and thermal evolution calculation
of a planet. 
\textit{Owen \& Wu (2013)} concluded that evaporation is the driving
force of evolution for close-in Kepler exoplanets.
It is therefore plausible to assume that some close-in exoplanets may be 
surrounded by a circum-planetary material.
(3) Strong interaction with the host star forces the planet into 
synchronization and circularization of its 
rotation and orbit (Zahn 1977; Tassoul \& Tassoul 1992; 
Bodenheimer, Lin \& Mardling 2001). (4) Strong interaction with the host star, 
and/or presence of an additional
planet may lead to variations in the orbital period of the planet
(Agol et al. 2005; Holman \& Murray 2005; Ford et al. 2011, 2012; 
Steffen et al. 2012; Maciejewski et al. 2013; Mazeh et al. 2013). 

Kepler mission discovered thousands of new transiting extrasolar planet
candidates (Borucki et al. 2011).
The unprecedented photometric precission of Kepler has made it possible to detect
transits by Earth size planets (Fressin et al. 2012;
Borucki et al. 2012), planetary optical secondary eclipses and phase variations 
(see e.g. Jackson et al. 2014; Sanchis-Ojeda et al. 2014; Esteves, 
De Mooij \& Jayawardhana 2013 and references therein), amongst other subtle phenomena.

A unique close-in Mercury-size Kepler exoplanet candidate KIC012557548b has been discovered
recently by \textit{Rappaport et al. (2012)}.
Unlike all other exoplanets it exhibits significant variability in 
the transit depth (Fig. 1). The shape of the transit is highly asymmetric, with 
a significant brightening just before the eclipse, sharp ingress 
followed by a smooth egress (Fig. 2). 
The light curve of this planet was studied in more detail
by \textit{Brogi et al. (2012)}, \textit{Budaj (2013)}, \textit{Kawahara et al. (2013)}, \textit{Croll et al. (2014)}, and \textit{van Werkhoven et al. (2014)}. 
The planet also has an extremely short orbital period of
0.65356(1) days (15.6854 hours). The host star that is apparently being occulted is KIC012557548, a V = 16
magnitude K-dwarf with $T_{eff} \simeq 4400$ K.
\textit{Rappaport et al. (2012)}
suggested that the planet has size not larger than Mercury, and is
slowly disintegrating/evaporating, creating a comet-like tail.
\textit{Perez-Becker \& Chiang (2013)} proposed a model of the atmosheric 
escape via the thermal wind. It is effective only for planets which
are less massive than Mercury. Gravity of the more massive planets 
would provide too deep potential barier for the wind.
Another close-in Kepler exoplanet candidate KIC8639908b
($R_{p} \leq 1.06 R_{\oplus}$), detected recently
also by \textit{Rappaport et al. (2014)} with an orbital period of
0.910022(5) days, exhibits a distinctly asymmetric transit profile, likely indicative 
of the emission of dusty effluents, and reminiscent of KIC012557548b. The host
star has $T_{eff} \simeq 4435$ K, it is a V = 15.9 magnitude K-dwarf.
Mass loss and possible comet-like tail was detected also in GJ 436b in Ly$\alpha$ by \textit{Kulow et al. (2014)}.

\begin{figure}
\centering
\includegraphics[width=65mm]{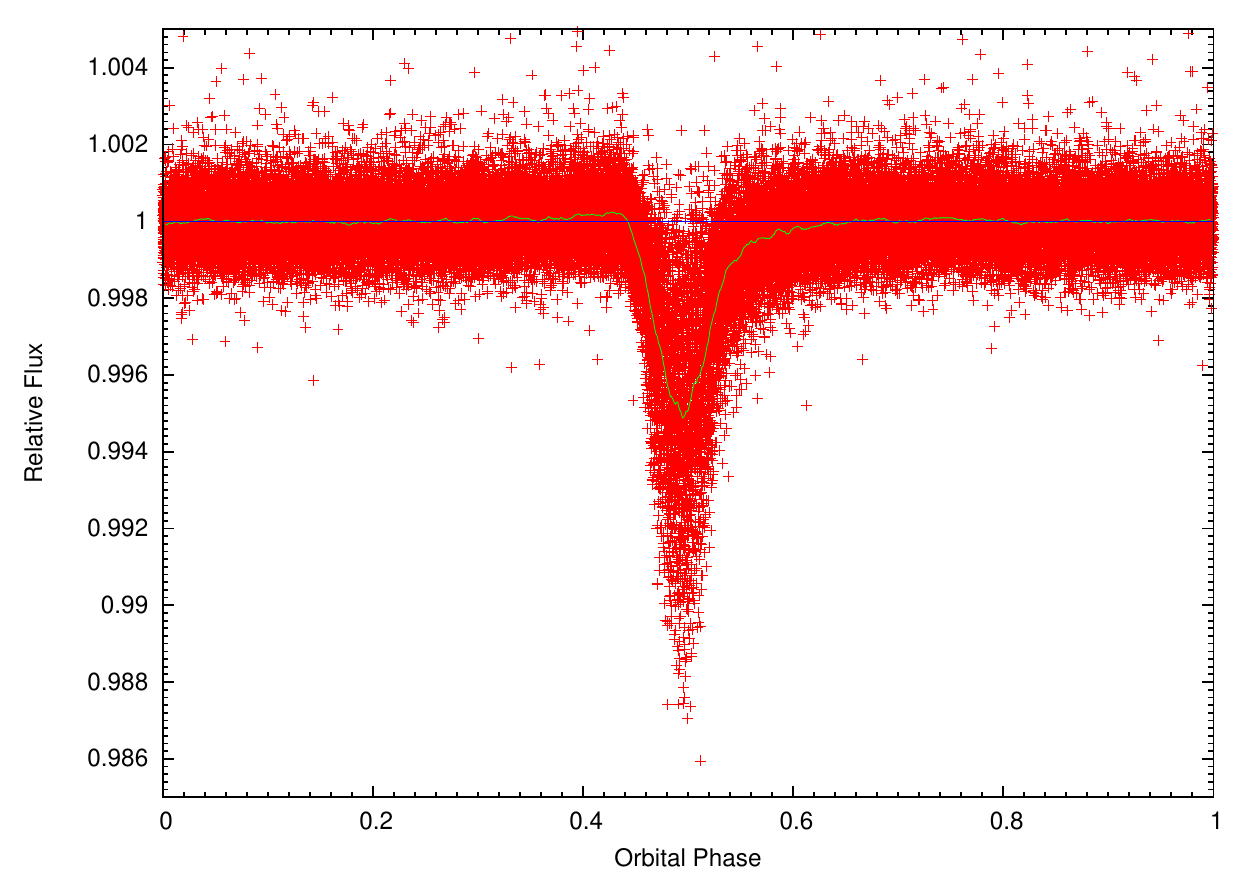} 
\caption{The transit light curve of KIC012557548b with the transit (Borucki et al. 2011; analyzed by Budaj 2013). It is best represented by an exoplanet with a comet-like tail.}
\label{Fig. 1}
\end{figure}

In this paper, we search for evidence of disintegration in the shortest period Kepler exoplanet candidates, by examining their transit light curves for asymmetric features. 
The layout of the paper is as follows.
In \textit{Section 2}, we provide an overview
of our exoplanet sample. In \textit{Section 3}, we describe the data analysis
and in \textit{Section 4} our motivation. 
\textit{Section 5} is the main part of our work. It describes obtained results. Our findings
are summarized in \textit{Section 6}.
   

\section{Motivation, aims and sample selection}
Our main aim is to search for comet-like tails similar to KIC012557548b and 
for long-term orbital period variations. 
We are curious about frequency of comet-like tail formation among short-period Kepler exoplanet candidates.
Close-in exoplanets, like KIC012557548b, are subjected to the greatest planet-star interactions. 
We therefore concentrate on the short-period exoplanet
candidates with periods similar to KIC012557548b.
Kepler mission exoplanet candidates are described in the catalog of 
\textit{Batalha et al. (2013)}. From this catalog we chose a sample of 
20 candidates with the shortest orbital period.
An overview of our sample is in \textit{Table 1}.
This sample covers the orbital periods in the range of
0.370 to 0.708 days, and stellar effective temperatures in the range of 
$T_{eff} \simeq 3900 - 6600$ K. Consequently, the incident flux hitting the exoplanet may not be exactly 
the same as in the case of KIC012557548b, but  
may slightly vary by about 1-2 orders of magnitude.

\begin{table}[!h]
\centering
\caption{An overview of our sample (Batalha et al.
2013). This sample covers the orbital periods in the range of 0.370 to 0.708 days, and stellar effective temperatures in the range of $T_{eff} \simeq 3900 - 6600$ K.}
\label{Table 1}
\begin{tabular}{ccc}\hline
\hline
KIC Number & Orbital period & $T_{eff}$\\ 
& (days) & (K)\\
\hline
\hline
3848972 & 0.3705290    & 5286\\
8561063 & 0.4532875    & 4188\\
6666233 & 0.5124040    & 3932\\
6047498 & 0.5187486    & 5359\\
9030447 & 0.5667281    & 6693\\
6934291 & 0.5678562    & 4927\\
4055304 & 0.5710391    & 5022\\
10024051 & 0.5773752   & 5062\\
8235924 & 0.5879933    & 4003\\
11774303 & 0.6140747  & 6271\\
10975146 & 0.6313298  & 4369\\
10028535 & 0.6630983  & 4994\\
10468885 & 0.6640818  & 5013\\
11600889 & 0.6693163  & 5627\\
8278371 & 0.6773911   & 5731\\
5513012 & 0.6793614   & 5375\\
10319385 & 0.6892040  & 5719\\
9761199 & 0.6920069   & 4060\\
9473078 & 0.6938521   & 5353\\
5972334 & 0.7085982   & 5495\\
\hline
\end{tabular}
\end{table}


\section{Light curve analysis and model fitting}
We used the publicly available 17 quarter Kepler data  
in the form of Pre-search Data Conditioning Simple Aperture Photometry (PDCSAP) fluxes.
Only the long cadence data were used, as they are sufficient in revealing the large-scale asymmetries caused by 
circum-planetary material.  
Kepler observations were treated and analysed similarly as in 
\textit{Budaj (2013)}.  
There is an offset between Kepler fluxes from different quarters.
Consequently, fluxes within each quarter were normalized to unity.
We then improved the preliminary orbital period of the exoplanet (Batalha et al.
2013) using the method of phase
dispersion minimization (PDM) described later.
The data were cut into segments each covering one orbital period.
Each segment of data was fitted with a linear function.
During the fitting procedure the part of the data covering the transit
was excluded from the fit. Consequently, the linear trend was removed
from each chunk of data (including the transit data).
The final value of the orbital period was then found in this detrended
data.
Finally the data were phased with this new orbital period. We used phase 0.5 for transits.
This method can effectively remove the long term variability
(mainly variability of the host star due to spots and rotation) 
while it does not introduce any nonlinear trend to the phase 
light curve.
To reduce the noise, the phased light curve was subject to a running 
window averaging. We used a window with the width of 0.01
and step of 0.001 (in units of phase) in all cases. 

We searched for the signatures of a circum-planetary material
in these light curves. 
For this purpose the final transit light curve of each planet was
fitted with a theoretical light curve (see left panels in appendix A), and the residuals were examined
for abnormalities (see mid panels in appendix A). We employed the \textit{Mandel \& Agol (2002)}
transit model with the free parameters: mid transit time $T_c$ (in TDB-based BKJD), period $P$ (days),
planet-to-star radius ratio $R_p/R_s$, normalised semi-major axis $(R_p+R_s)/a$,
line-of-sight inclination $i$ (deg), and quadratic limb darkening parameters
$q_1=(u_1+u_2)^2$ and $q_2=0.5 u_1 (u_1+u_2)^{-1}$
parameterised as per \textit{Kipping (2013)}. We first compute the
light curve for a single transit epoch, sampled at 0.0005 in phase ($<1$
minute sampling for a 1 day period candidate). The model light curve
is then convolved with a box-car with width of 30 minutes, simulating
the integration time of a long-cadence exposure. This template
transit model is then interpolated using a B-spline, and evaluated at 
the time stamps of the observed light curve to arrive at the final   
model. To speed up the computation
process, we reduce the number of points by selecting only the
in-transit and 0.1 phase of the out-of-transit portions of the light curve for the fitting.
This cropping of the light curve is justified since we are interested in only the distortions
to the transit shape, not the out-of-transit variations. To find the best fit parameters and
explore the degeneracies and uncertainties, we perform a Markov Chain Monte
Carlo (MCMC) minimisation using the \emph{emcee} ensemble sampler
(Foreman-Mackey et al. 2013). To better account for sources of
errors in the photometry, individual measurement errors of 
the light curves are inflated such that the reduced $\chi^2$ is at unity before
the MCMC routine, enabling a more realistic estimate the posterior probability
distribution. The best fit parameters and their errors are derived
from the marginalised posterior probability distributions after the MCMC
analysis. 

We also searched for possible long-term changes of the orbital period. 
First we improved the preliminary orbital period of the extrasolar 
planet using the method of phase dispersion minimization 
(PDM -- Stellingwerf 1978; application PDM2 ver. 4.13\footnote{Actual version of the software package
is available on the webpage http://www.stellingwerf.com} -- Stellingwerf 2004) 
and then using the Fourier analysis (FA). 
The detrended data were used for this purpose (as described earlier
in this section).
We assumed that the period changes linearly:
\begin{equation}
  \label{equation1}
  P=P_{\mathrm{0}}+\beta t
\end{equation}
where $\beta$ is a dimensionless value, but is often
expressed in days/million years (d/Myr). The output from the analysis 
(PDM2 4.13 in this case) is a curve (see right panels in appendix A), 
which shows the dependence of $\Theta_{min}$ as a function of $\beta$. 
$\Theta_{min}$ is a dimensionless statistical parameter 
(Stellingwerf 1978). The minimum value of $\Theta_{min}$ indicates 
the value of the period change -- $\beta$. 

For candidates where the PDM analysis indicated potential period variations, 
we broke the light curve into segments of 2000/4000 points, and re-performed the
MCMC analysis. The minimisation is performed globally such that all
the light curve segments are fitted simultaneously, allowing individual $T_c$ values 
for each segment of the light curve, but shared transit geometry parameters 
$R_p/R_s$, $(R_p+R_s)/a$, and $i$. We fixed $P$ to the best fit period from the initial
analysis. The fitted $T_c$ values are then inspected for non-linear variations 
with respect to the centre time of each light curve segment.  


\section{Exoplanet candidate KIC012557548b}
To test the effectiveness of our techniques, we first applied
the procedure, described above, to this exoplanet candidate.

\begin{figure}[!h]
\centering
\includegraphics[width=65mm]{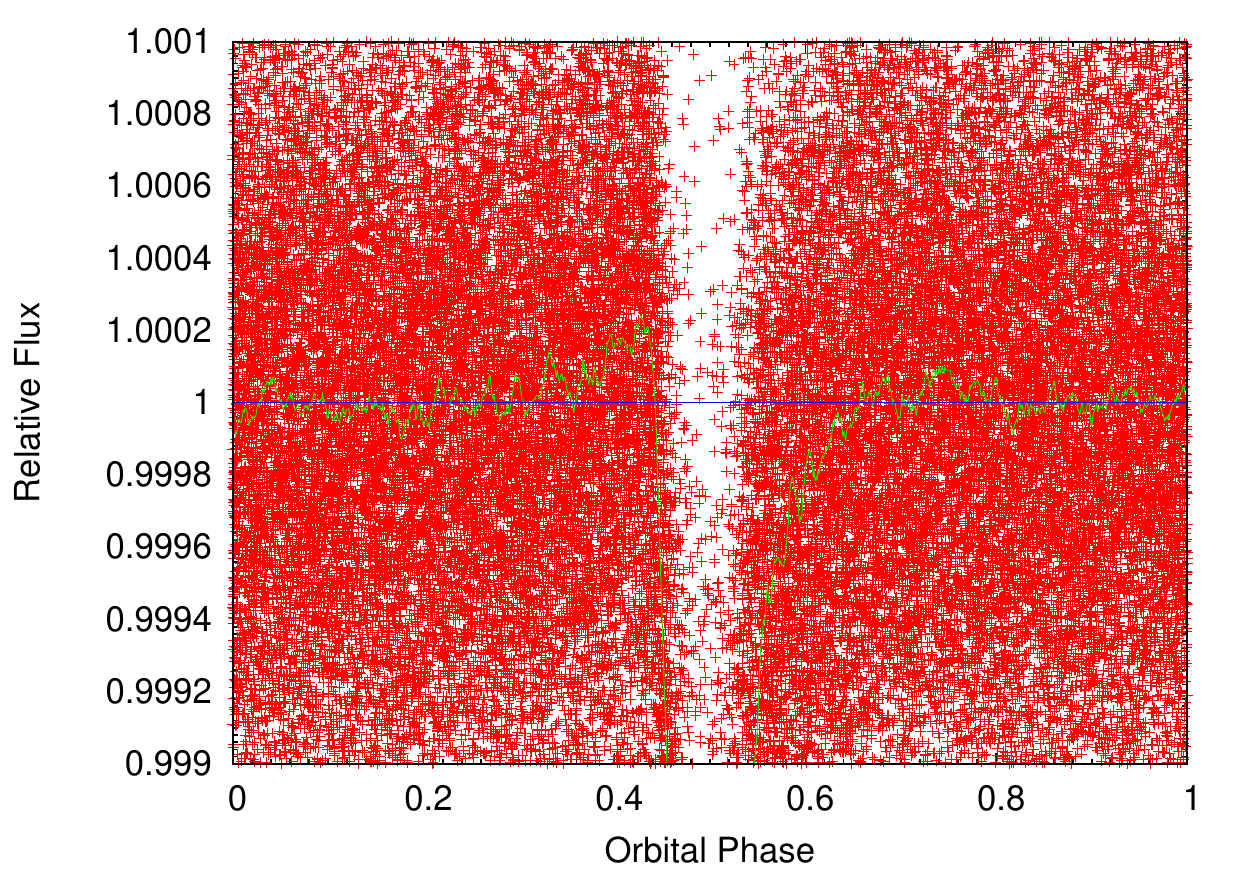}
\caption{The transit light curve of KIC012557548b (Borucki et al. 2011; analyzed by Budaj 2013), cropped
to enhance the asymmetric features. It shows a significant 
pre-transit brightening, sharp ingress followed by a short sharp egress and
long smooth egress, and a weak post-transit brightening.}
\label{Fig. 2}
\end{figure}

KIC012557548b can be used as an extreme example of the exoplanet with 
a circum-planetary material. It was discovered by \textit{Rappaport et al. (2012)}.
The light curve of the Kepler exoplanet candidate KIC012557548b is 
peculiar and very interesting.
It shows a significant brightening just
before the eclipse -- pre-transit brightening, 
sharp ingress followed by a short sharp egress and long smooth egress,
and a weak post-transit brightening (Brogi et al. 2012; Budaj 2013).
A close-up of the transit features are shown in \textit{Fig. 2}. 

Moreover, the candidate exhibits strong variability in the transit core on timescale 
of one day (Rappaport et al. 2012), and variability
in the egress on the timescale of about 1.3 years (Budaj 2013).
\textit{Rappaport et al. (2012)} suggest that KIC012557548b is 
a slowly disintegrating/evaporating planet what creates a comet-like tail. 
\textit{Brogi et al. (2012)} and \textit{Budaj (2013)} reanalyzed 
the light curve in detail and
both validated the disintegrating-planet scenario by modeling. Both
brightenings are caused by the forward scattering on dust particles 
in the tail, which have typical radii of about 0.1-1 micron. 
Strong variability in the transit depth is a consequence of changes 
in the cloud optical depth.

\begin{figure}[!t]
\centering
\includegraphics[width=80mm]{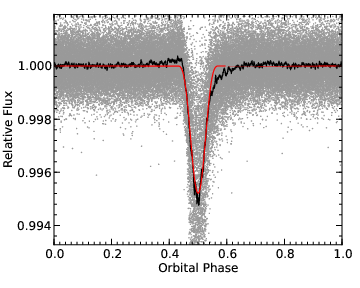}
\caption{The transit light curve of KIC012557548b with the
transit (Borucki et al. 2011; analyzed by Budaj 2013). 
It was fitted with a theoretical light curve 
similarly as light curves in \textit{appendix A} 
(see Section 3 and left panels in appendix A).}
\label{Fig. 2}
\end{figure}

\begin{figure}[!t]
\centering
\includegraphics[width=80mm]{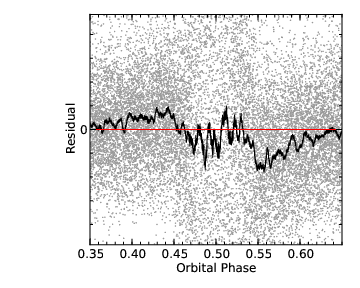}
\caption{The transit residuals of KIC012557548b obtained
similarly as transit residuals in \textit{appendix A} 
(see Section 3 and mid panels in appendix A). 
A standard Mandel-Agol transit fits this candidate poorly.
We can see the pre-transit brightening between
$0.39 < P < 0.45$, as well as the smooth egress between $0.55 < P < 0.65$.
Large scatter observed between $0.45 < P < 0.55$ 
indicate the variability in the transit depth.}
\label{Fig. 3}
\end{figure}  

Subsequently, we fitted the observed light curve with the theoretical model light curve,
assuming a spherical planet (Fig. 3), and obtained transit 
residuals of KIC012557548b (Fig. 4) as per \textit{Section 3}. 
The best fit quadratic limb darkening coefficients were
$q_1 = 0.98(8)$ and $q_2 = 0.98(7)$. The best fit system parameters are: mid transit time $T_c=0.0404(2)$ (BKJD), period $P=0.65355$ days, 
planet-to-star radius ratio $R_p/R_s=0.3(1)$, normalised semi-major axis $(R_p+R_s)/a=0.60(5)$, and 
line-of-sight inclination $i=58(4)$ (deg).    
\textit{Fig. 4} shows a significant fluctuation in these residuals.
This signature of the residuals indicate the circum-planetary material.
The residuals also reveal the pre-transit brightening between
$0.39 < P < 0.45$, as well as the smooth egress between $0.55 < P < 0.65$.
We can see the pre-transit brightening as a positive fluctuation, and the smooth egress
as a negative fluctuation. Large scatter observed between $0.45 < P < 0.55$ 
indicate the variability in the transit depth. 
These additional features of transit residuals suggest the comet-like tail.

\begin{figure}[!h]
\centering
\includegraphics[width=65mm]{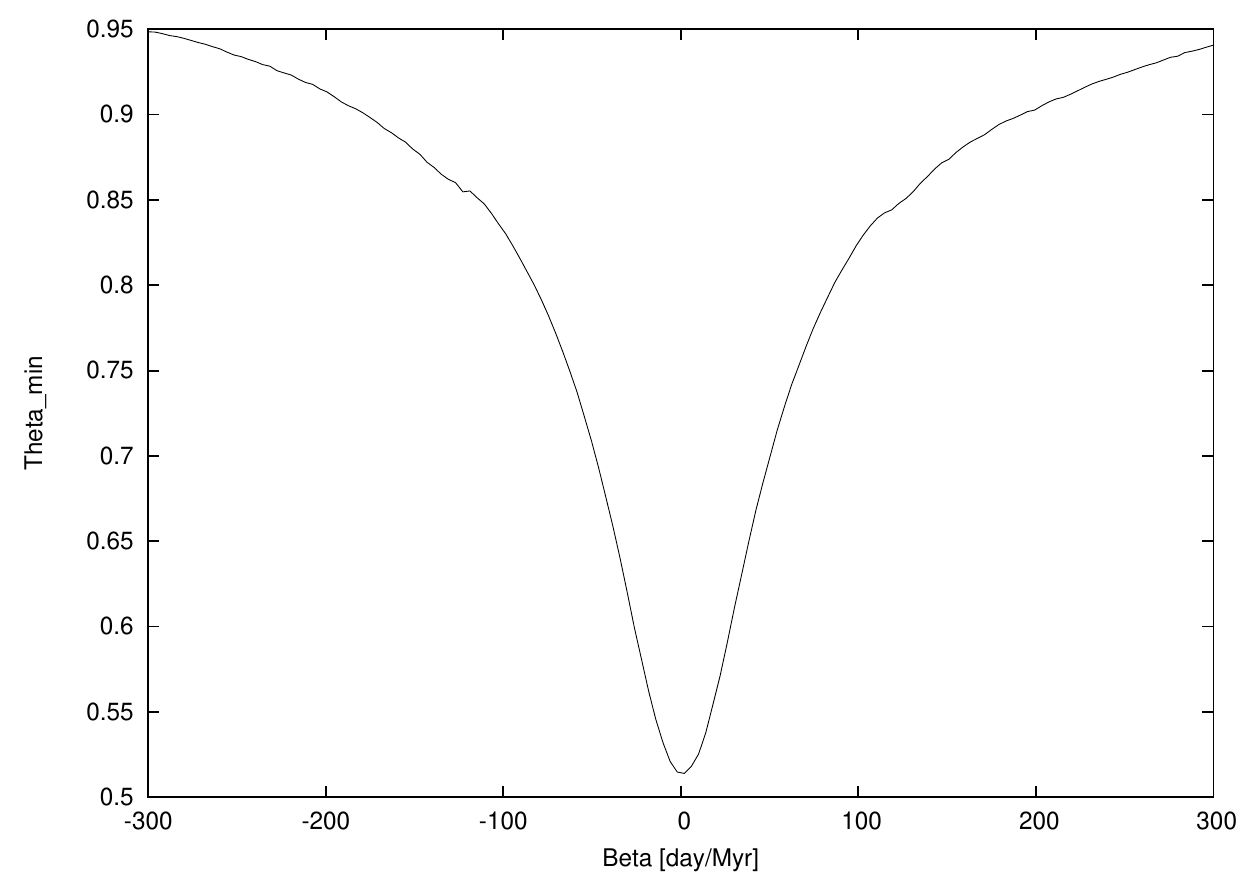}
\caption{Serach for a long-term change of the orbital period of KIC012557548b based on 17 quarter data.
Figure shows that there is no significant evidence for the long-term period change.}
\label{Fig. A15}
\end{figure}

We then searched for evidence of transit-timing variations in KIC012557548b using the PDM method.
Since KIC012557548b is close-in exoplanet candidate that is apparently
being disintegrated and losing material, one might expect
all kinds of interaction that could lead to the long-term period
variation. That is why \textit{Budaj (2013)} also searched for 
possible long-term changes of the orbital period with the PDM method 
(Stellingwerf 1978; application PDM2 4.13 -- Stellingwerf 2004). 
He obtained $\beta=0.3\pm0.5$ d/Myr, which means that there is no 
significant evidence for the long-term period change. 
Since \textit{Budaj (2013)} searched for possible long-term changes 
of the orbital period based on 14 quarters of Kepler data, we repeated this analysis using 17 
quarters of data. We can confirm the author's result that there is no significant evidence
for the long-term orbital period variation: $\beta=0.9 \pm 1.1$ d/Myr (Fig. 5).                   
We also improved the orbital period of the exoplanet KIC012557548b and obtained $P=0.65355390$ days using the PDM
and $P=0.65355380$ days using the FA method. These results are very similar to the fit period result. These values are also in the good agreement with the values reported in \textit{Rappaport et al. (2012)}, \textit{Budaj (2013)}, and \textit{van Werkhoven et al. (2014).}


\section{Results}
The following subsections include a description of the candidates, formated as per
our discussion of KIC012557548b. Each subsection includes results from both 
(1) analysis of the transit residuals, where we searched for 
comet-like tails similar to KIC012557548b, and (2) analysis of the orbital
periods, where we searched for long-term orbital period variations.
Interesting exoplanet candidates are described in individual 
subsections. The last subsection includes the exoplanets that 
do not show any significant features.


\subsection{Exoplanet candidate KIC3848972}
The Kepler exoplanet candidate KIC3848972 is interesting for
several reasons.
Although the transit light curve (Fig. A1 -- see left panel) does not show signs of a pre-transit
brightening, post-transit brightening, nor significant variability in 
the transit depth, an other hand its transit is significantly deeper (0.002) then
transits of other planets from our sample, and it has a V-type shape, which might indicate a grazing
eclipsing binary.
Subsequently, we fitted the light curve (Fig. A1 -- see left panel) and obtained transit residuals of KIC3848972 
(Fig. A1 -- see mid panel). The transit residuals of KIC3848972 show only a linear downward trend from sine-like background variability. We obtained the best fit quadratic limb darkening coefficients of $q_1 = 0.9(1)$ and $q_2 = 0.9(3)$ for the star. The best fit system parameters are: mid transit time $T_c=534.34045(5)$ (BKJD), period $P=0.37053$ days, 
planet-to-star radius ratio $R_p/R_s=0.18(8)$, normalised semi-major axis
$(R_p+R_s)/a=0.71(4)$, and 
line-of-sight inclination $i=49(6)$ (deg). The parameter $R_p/R_s=0.0435$, presented by \textit{Batalha et al. 2013}, is very different from what we got for $R_p/R_s$, due to the degeneracy between $R_p/R_s$, inclination $i$, and $(R_p+R_s)/a$.
Indeed \textit{Col\'on, Ford \& Morehead (2012)}, based on a significant color
change during the transit event, identify KIC3848972 as 
a false positive, which may consist of an evolved giant star that is 
redder and several magnitudes brighter than the eclipsing star. 
\textit{Slawson et al. (2011)} found an orbital period for this system that is
twice as long. Moreower, we found a small difference between the odd and even
transits in 
this system (0.000140) and a weak periodic background variability, which is very similar to an RS Canum Venaticorum-like distortion wave (see e.g. Luddington 1978). That is why we consider the eclipsing binary alternative as possible. We also observed a similar peculiarity at KIC9761199 (see subsection 5.7). On \textit{Fig. 6} we can see that the sine-like distortion wave is extends throughout the double-period phase folded light curve. The distortion wave is slightly shifting in phase during the Kepler observations, but we always observed only one sine-wave. The background variations may potentially be due to spot modulation from one of the stars.

\begin{figure}[!h]
\centering
\includegraphics[width=80mm]{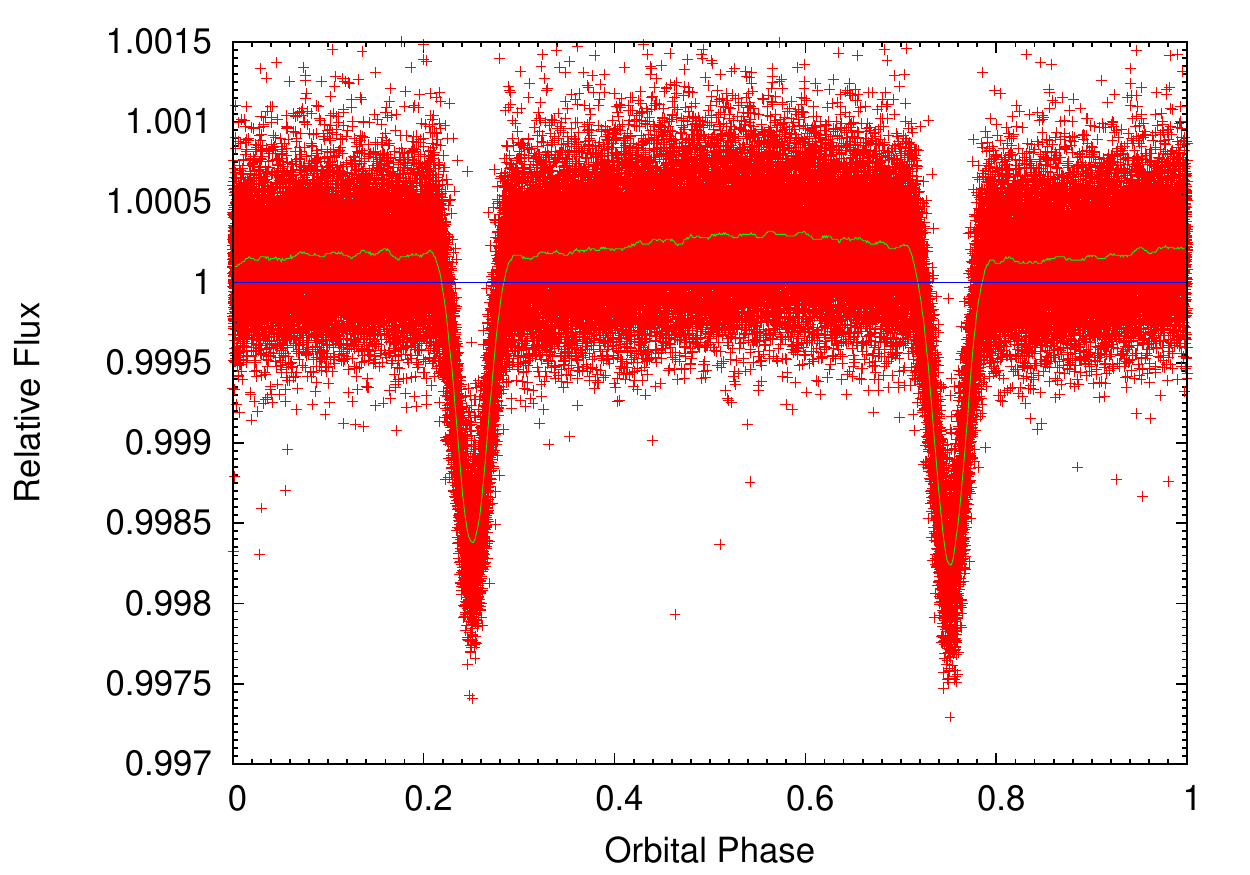}
\caption{The double-period phase-folded light curve of KIC3848972. The system exhibits a small difference between the odd and even transits, and a sinusoidal background variability.}
\label{Fig. 4}
\end{figure}

Next we searched for long-term orbital period changes. 
First we improved the orbital period. We obtained period $P=0.37052870$ 
days using the PDM and $P=0.37052750$ days using the FA method, which are very similar to the fit period result. 
These values are also in the good agreement with the value 
$P=0.3705290$ days, reported in \textit{Batalha et al. (2013)}. However, based on our period analysis, the two times alias orbital period is also possible. If the period
is indeed two times longer, both stars would have very similar temperatures.   
\textit{Fig. A1 -- right panel} shows that there is no significant
evidence for the long-term period change ($\beta=-0.8 \pm 1.4$ d/Myr). 


\subsection{Exoplanet candidate KIC8561063}
The Kepler exoplanet candidate KIC8561063 is similar to the previous
exoplanet candicate. It shows relatively deep transit (0.0015)
with a V-type shape, which needs to be checked, since it
may be indicative of an eclipsing binary (Fig. A2 -- see left panel). An other hand, we do not find significant difference between the odd and even transits in 
this system. The light curve also does not show any signs of a pre-transit
brightening, post-transit brightening, nor significant variability in 
the transit depth.
We modelled the transit light curve and obtained transit residuals from 0.35 to 0.65 in phase units. They are depicted in \textit{Fig. A2 -- mid panel}, and also do not show any signs of a comet-like tail. The fitting derived model has quadratic limb darkening coefficients of 
$q_1 = 0.0001(8)$ and $q_2 = 0.7(1)$.
The best fit system parameters are: mid transit time $T_c=488.7930(1)$ (BKJD), period $P=0.45329$ days, 
planet-to-star radius ratio $R_p/R_s=0.044(4)$, normalised semi-major axis $(R_p+R_s)/a=0.2(1)$, and 
line-of-sight inclination $i=76(2)$ (deg). 

Subsequently, we improved the orbital period and obtained 
$P=0.45328705$ days using the PDM and $P=0.45328830$ days using the FA
method, which are very similar to the fit period result. However, based on our period analysis, the two times alias orbital period is also possible. If the period
is indeed two times longer, both stars would have the same temperatures. \textit{Fig. A2 -- right panel} shows that there is 
no significant evidence for the long-term period evolution ($\beta=0.92 \pm 0.93$ d/Myr).


\subsection{Exoplanet candidate KIC6666233}
Its transit (Fig. A3 -- left panel) is shallower (0.00037) then transits 
mentioned above, and has an U-shape, which is typical for exoplanet 
transits.
At the same, we detected a small difference between the odd and even 
transits (0.000054; see Fig. 7), potentially indicative of an eclipsing binary system with twice the orbital
period.   

Subsequently, we modelled the transit light curve (Fig. A3 -- see left panel). The best fit limb darkening coefficients 
of the star were $q_1 = 0.26(4)$ and $q_2 = 0.8(1)$. The best fit system parameters are: mid transit time $T_c=546.2643(5)$ (BKJD), period $P=0.51241$ days, planet-to-star radius ratio $R_p/R_s=0.016(2)$, normalised semi-major axis $(R_p+R_s)/a=0.27(1)$, and 
line-of-sight inclination $i=85(4)$ (deg).  
The light curve and transit residuals (Fig. A3 -- left and mid 
panels) do not show any indication of the circum-planetary material.

\begin{figure}[!h]
\centering
\includegraphics[width=80mm]{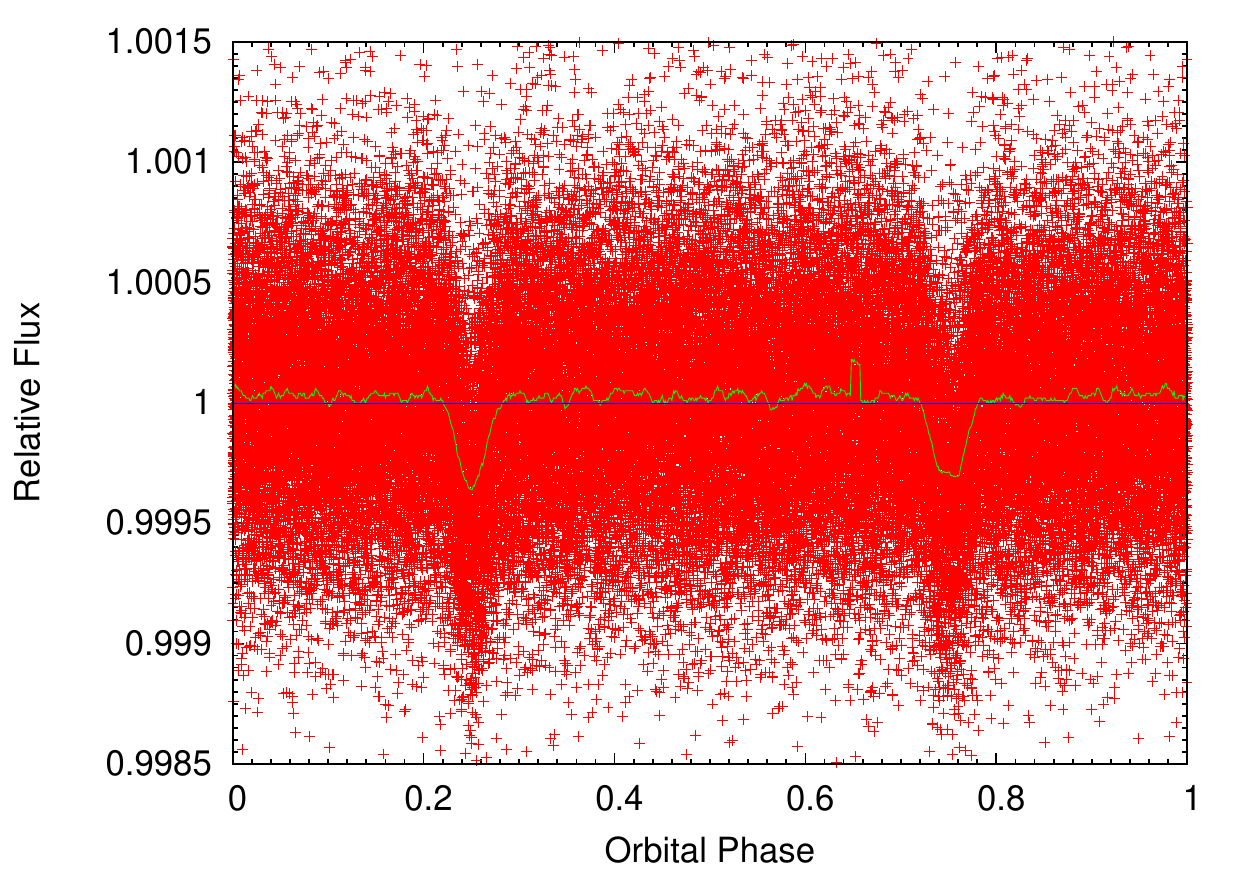}
\caption{The double-period phase-folded light curve of KIC6666233 with
a small depth difference between the odd and even transits in this system.}
\label{Fig. 5}
\end{figure}

We improved the orbital period and obtained $P=0.51240883$ days using the PDM and $P=0.51240860$ days using the FA method. 
These values are in the good agreement with the value 
$P=0.5124040$ days, reported in \textit{Batalha et al. (2013)}. However, based on our period analysis, the two times alias orbital period is still possible.  
There is no significant evidence for the long-term period change ($\beta=-1.0 \pm 2.3$ d/Myr).   


\subsection{Exoplanet candidate KIC6047498}
The light curve of
this candidate (Fig. A4 -- left panel) exhibits a V-shaped transit with depth of 0.0007, showing  
no pre-transit brightening, post-transit brightening, nor
significant variability in the transit depth. 
The transit residuals also do not show any signs of a comet-like tail (Fig. A4 -- see mid panel). The best fit quadratic limb darkening coefficients were $q_1 = 0.94(5)$ and $q_2 = 0.95(4)$. The best fit system parameters are: mid transit time $T_c=534.1331(3)$ (BKJD), period $P=0.51873$ days, planet-to-star radius ratio $R_p/R_s=0.3(2)$, normalised semi-major axis $(R_p+R_s)/a=0.76(5)$, and 
line-of-sight inclination $i=43(5)$ (deg). The parameter $R_p/R_s=0.02919$, presented by \textit{Batalha et al. (2013)}, is very different from our derived $R_p/R_s$, due to the degeneracy between $R_p/R_s$, inclination $i$, and $(R_p+R_s)/a$.
We also detected a small difference between
the odd and even transits in this system (0.000135; see Fig. 8), which is similar
to the difference measured at the candidate KIC3848972.

\begin{figure}[!h]
\centering
\includegraphics[width=80mm]{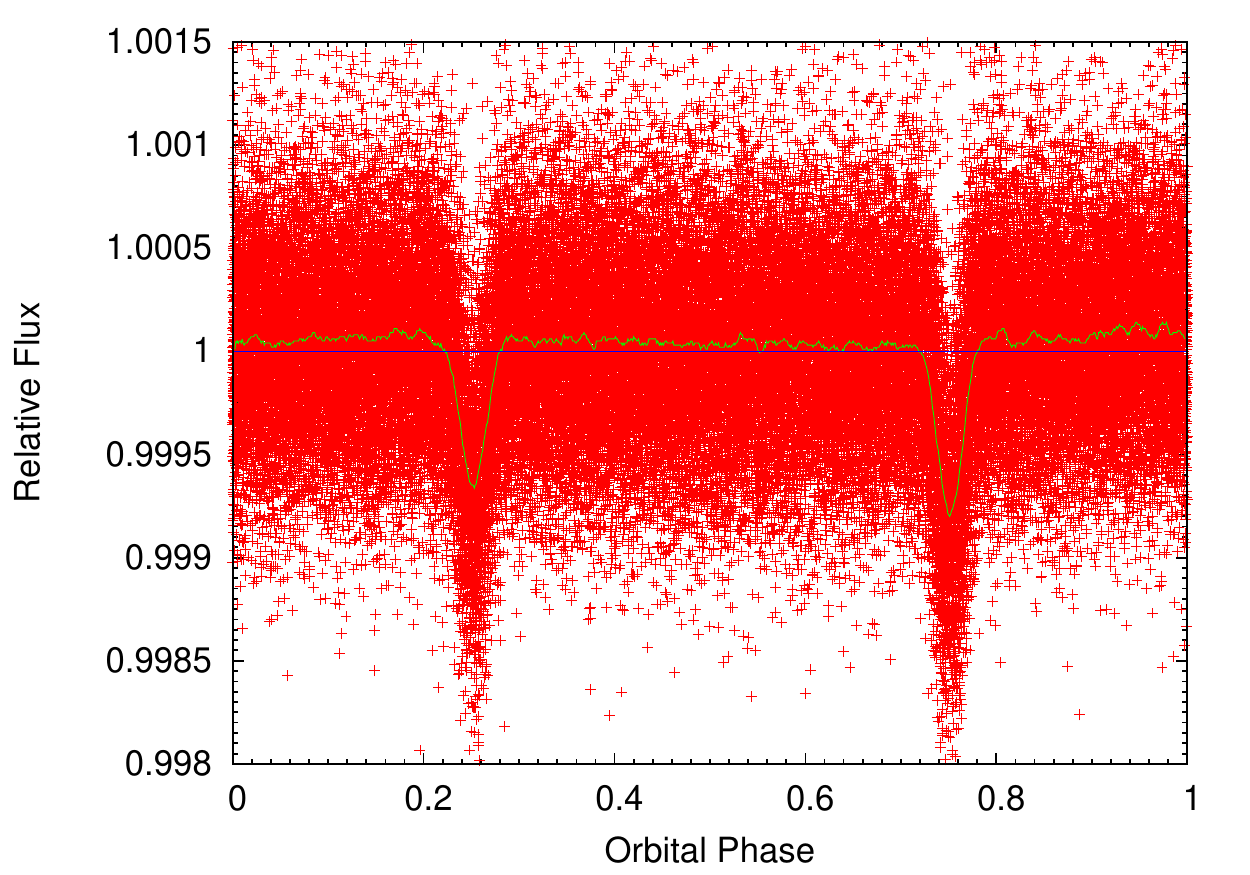}
\caption{The double-period phase-folded light curve of KIC6047498 with
a small difference between the odd and even transits in this system.}
\label{Fig. 8}
\end{figure}  

\begin{figure}[!h]
\centering
\includegraphics[width=80mm]{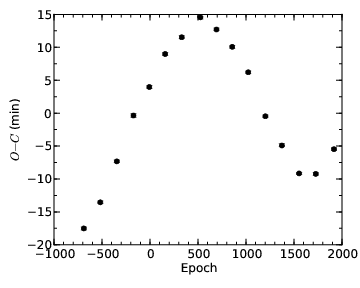}
\caption{The $T_c$-free MCMC analysis in case of KIC6047498. Figure shows the O-C diagram of mid transit time variations. The O-C exhibits a semi-periodic TTV variability, indicative of an outer companion with an orbital period of $ > 2000$ days.}
\label{Fig. 8}
\end{figure}  

Further we searched also for a long-term orbital period change. First we
improved the orbital period. We obtained period $P=0.51872845$ days 
using the PDM and $P=0.51872930$ days using the FA method, which are in the good agreement with the fit period result, and also with the value presented by \textit{Batalha et al. (2013)}, with the possibility of the two times alias. 
\textit{Fig. A4 -- right panel}
shows that there is possible evidence for the long-term period change,
$\beta=-15.05017 \pm 1.94760$ d/Myr, that means a potential shortening of the orbital period. Since the PDM analysis indicated a potential period variation, 
we broke the light curve into segments of 4000 points in this case, and re-performed the
$T_c$-free MCMC analysis. This analysis confirmed our results from the PDM analysis about long-term period change. We found a semi-periodic TTV signal associated with the transit, indicative of an outer companion with an orbital period of $ > 2000$ days (Figs. 9 and 10).    

\begin{figure}[!h]
\centering
\includegraphics[width=75mm]{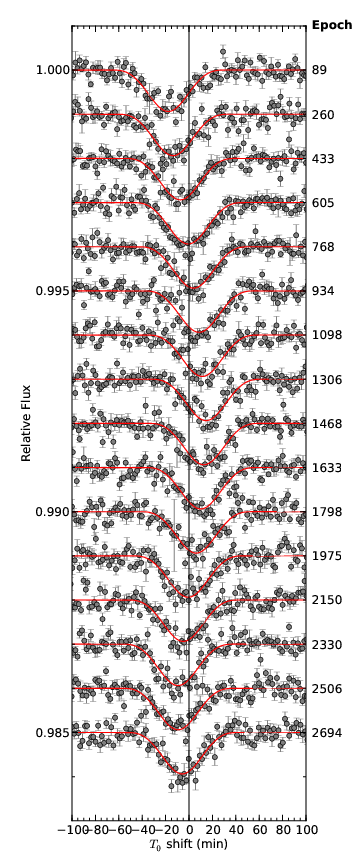}
\caption{The transit for each segment of the KIC6047498 light curve,
with best fit model overplotted. The mean transit epochs for each segment are labelled on the left. 
Each segment is arbitrarily shifted vertically for clarity. The solid vertical line shows the mean $T_c$ value. 
The transit centres are visibly shifted from segment to segment.}
\label{Fig. 8}
\end{figure}  


\subsection{Exoplanet candidate KIC9030447}
The light curve of the exoplanet candidate KIC9030447 
is peculiar (Fig. A5 -- see left panel).
It shows a small brightening, approximately half a phase from the transit, and a secondary eclipse-like signal approximately at phase 0.85. Subsequently, we obtained the model light curve and transit residuals of the exoplanet candidate. The best model fit did not describe the transit well. The best fit quadratic limb darkening coefficients 
of the star were $q_1 = 0.47(2)$ and $q_2 = 0.99(2)$. The best fit system parameters are: mid transit time $T_c=534.478(1)$ (BKJD), period $P=0.56677$ days, 
planet-to-star radius ratio $R_p/R_s=0.0091(1)$, normalised semi-major axis $(R_p+R_s)/a=1.000(8)$, and 
line-of-sight inclination $i=85(1)$ (deg).
The transit residuals of this exoplanet candidate reflect the peculiarities of the light curve, and residual fluctuations due to the poor fit (see Fig. A5 -- mid panel).
\textit{Ofir \& Dreizler (2013)} proposed that
such behavior is caused by pulsations, rather than transits or eclipses.   
Further we discovered that these additional features of the light curve slighly shifted in phase during the Kepler observations. Moreover, the secondary eclipse-like signal is changing its depth. That is why we consider the eclipsing binary on eccentric orbit alternative as more possible.

Subsequently, we improved the orbital period and obtained 
$P=0.56670873$ days using the PDM and $P=0.56680630$ days using
the FA method. \textit{Fig. A5 -- right panel} shows that there is 
no significant evidence for the long-term period change ($\beta=0.9 \pm 8.6$ d/Myr).


\subsection{Exoplanet candidate KIC11774303}
On first glance, the light curve of this exoplanet candidate
does not show any significant peculiarity. The transit residuals 
also do not show any convincing systematic fluctuations, and indicate 
that the light curve can be described well by the standard transit model
(Fig. A10 -- left and mid panels). The best fit quadratic limb darkening coefficients 
of the star were $q_1 = 0.118(9)$ and $q_2 = 0.7(2)$. The best fit system parameters are: mid transit time $T_c=457.3181(7)$ (BKJD), period $P=0.61408$ days, 
planet-to-star radius ratio $R_p/R_s=0.07(4)$, normalised semi-major axis $(R_p+R_s)/a=0.8(1)$, and 
line-of-sight inclination $i=37(11)$ (deg). 
However, \textit{Ofir \& Dreizler (2013)} detected significant 
differences between the odd and even transits in this system, and 
suggested that this candidate is an eclipsing binary with twice the
period. We also measured a transit depth difference (0.000228; Fig. 11) and we could confirm their conclusions
about transit differences. Based on this result we consider the eclipsing binary alternative as possible.

\begin{figure}[!h]
\centering
\includegraphics[width=80mm]{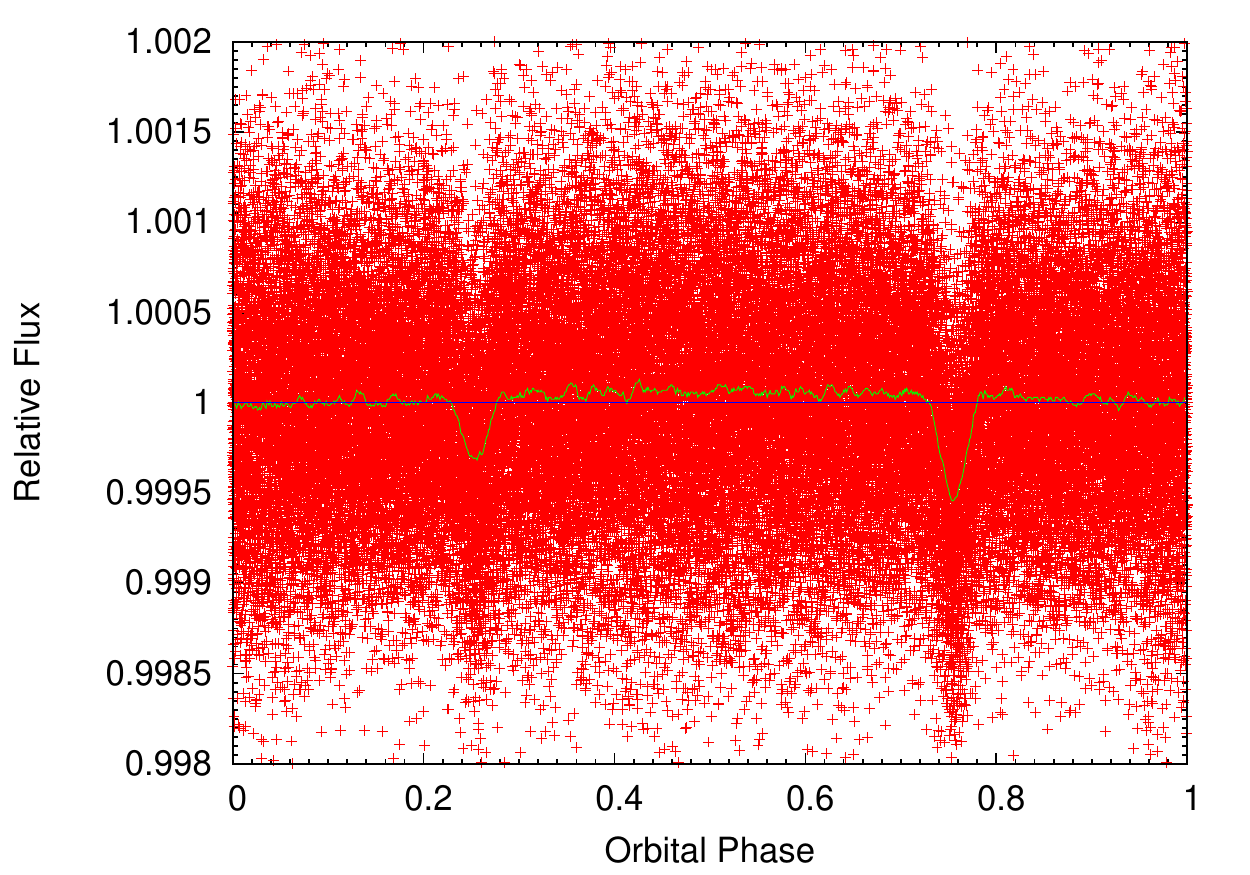}
\caption{The double-period phase-folded light curve of KIC11774303 with
a difference in depth between the odd and even transits in this system.}
\label{Fig. 7}
\end{figure}  

An other hand, we improved the orbital period and obtained period $P=0.61408403$ days 
using the PDM and $P=0.61409220$ days using the FA method, in agreement to the fit period result, but we 
could not confirm the two times period alias using the PDM/FA analysis. 
\textit{Fig. A10 -- right panel} shows that there is
no significant evidence for the long-term period change ($\beta=0.9 \pm 2.9$ d/Myr). 


\subsection{Exoplanet candidate KIC9761199}    
The Kepler exoplanet candidate KIC9761199 is also very interesting.
Its transit (Fig. A18 -- see left panel) is very deep (0.0045)
and has a V-type shape, which might indicate a grazing
eclipsing binary. 
The transit light curve does not show any signs of a pre-transit
brightening, post-transit brightening, nor significant variability in 
the transit depth.
Apart from that, the out-of-transit light curve exhibits a  
periodic background variability, which is very similar to an RS Canum Venaticorum-like distortion wave (see e.g. Luddington 1978). That is why we consider the eclipsing binary alternative as possible.  
The background variability is also very similar to the distortion wave observed by KIC3848972 (see subsection 5.1), but contrary to the KIC3848972, this wave is stronger. 
Subsequently, we fitted the light curve with a model light curve (Fig. A18 -- see
left panel) and obtained transit residuals of this
exoplanet candidate from 0.35 to 0.65 in phase units (Fig. A18 -- mid panel). The best fit quadratic limb
darkening coefficients were $q_1 = 0.94(5)$ and $q_2 = 0.96(1)$.
The best fit system parameters are: mid transit time $T_c=533.1669(1)$ (BKJD), period $P=0.69202$ days, 
planet-to-star radius ratio $R_p/R_s=0.22(9)$, normalised semi-major axis $(R_p+R_s)/a=0.43(3)$, and 
line-of-sight inclination $i=68(2)$ (deg).
The parameter $R_p/R_s=0.06943$, presented by \textit{Batalha et al. (2013)}, is very different from what we got for $R_p/R_s$, due to the degeneracy between $R_p/R_s$, inclination $i$, and $(R_p+R_s)/a$. 
The transit residuals show only a relatively stronger linear trend from sine-like background variability. 
Contrary to the KIC3848972, the linear trend in this case has an upward direction.
\textit{Ofir \& Dreizler (2013)} proposed that KIC9761199 is 
an eclipsing binary with twice the orbital period as suggested by 
\textit{Batalha et al. (2013)}. There is indeed a small
difference of 0.000704 between the odd and even transits
(see Fig. 12). On \textit{Fig. 12} we can see that the sine-like distortion wave is extends throughout the double-period phase folded light curve.   
Subsequently, we discovered that the sine-wave slightly shifted in phase during the Kepler observation. Sometimes we could observe double sine-wave and sometimes single sine-wave throughout the double-period phase folded light curve.

\begin{figure}[!h]
\centering
\includegraphics[width=80mm]{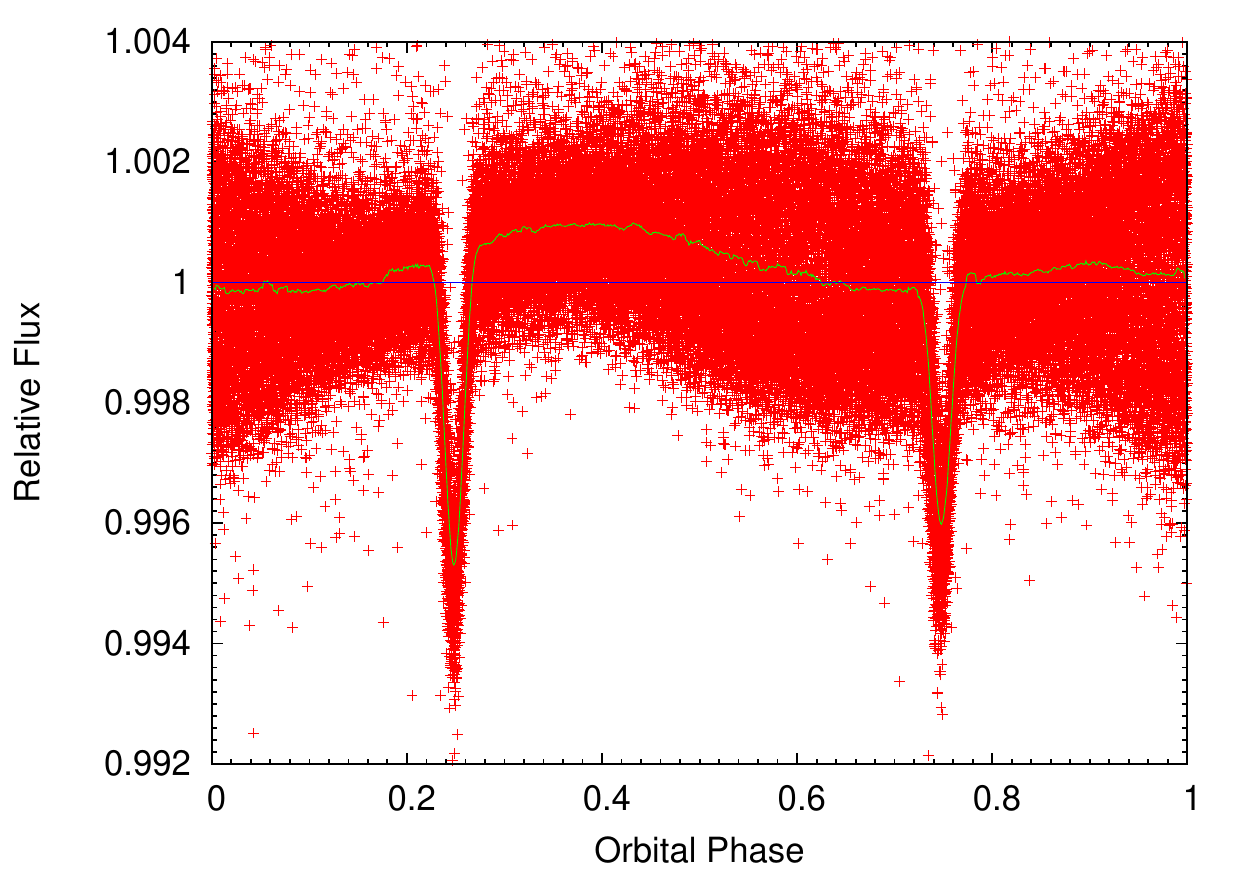}
\caption{The double-period phase-folded light curve of KIC9761199 with
a small difference between the odd and even transits in this system and with a sinusoidal background
variability.}
\label{Fig. 8}
\end{figure}

The background variability may potentially be due to the   
Doppler beaming effect, sometimes also called Doppler boosting,
induced by the stellar radial motion. This effect causes an increase 
(decrease) in the brightness of any light source approaching 
(receding from) the observer (e.g. Rybicki \& Lightman 1979; Loeb \& Gaudi
2003; Faigler \& Mazeh 2011). We did not see any evidence of ellipsoidal and reflection phase variations from this system. In addition, the sinusoidal variability shifted in phase during the Kepler observations, suggesting that it is not due to beaming. The background variations may potentially be due to stellar spot modulation from both of the stars when we observe double sine-wave, and from one of the stars, when we observe single sine-wave throughout the double-period phase folded light curve.      

Further we searched also for long-term orbital period change. First we
improved the orbital period. We obtained period $P=0.69201741$ days using 
the PDM and $P=0.69204530$ days using the FA method. These values are 
in the good agreement with the value $P=0.69202$ days obtained by the fitting procedure, and also in the good agreement with the value $P=0.6920069$ days reported 
by \textit{Batalha et al. (2013)}, but based on our period analysis, the two
times period alias could not be ruled out. \textit{Fig. A18 -- right panel} 
shows that there is a possible evidence for the long-term period change, with 
$\beta=-3.01003 \pm 0.43245$ d/Myr, which means a potential shortening of 
the orbital period. Since the PDM analysis indicates a potential period variation, 
we broke the light curve into segments of 2000 points in this case, and re-performed the
$T_c$-free MCMC analysis. This analysis confirmed our results from the PDM analysis about long-term period change, and revealed a periodic TTV signal with a period of $\sim 1500$ days, indicative of a possible massive outer companion (Figs. 13 and 14).

\begin{figure}[!h]
\centering
\includegraphics[width=80mm]{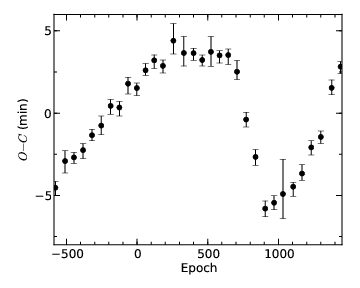}
\caption{The $T_c$-free MCMC analysis in case of KIC9761199. Figure shows the O-C diagram of mid transit time variations. We found a periodic TTV signal with a period of $\sim 1500$ days, indicative of a possible massive outer companion.}
\label{Fig. 13}
\end{figure}  

\begin{figure}[!h]
\centering
\includegraphics[width=75mm]{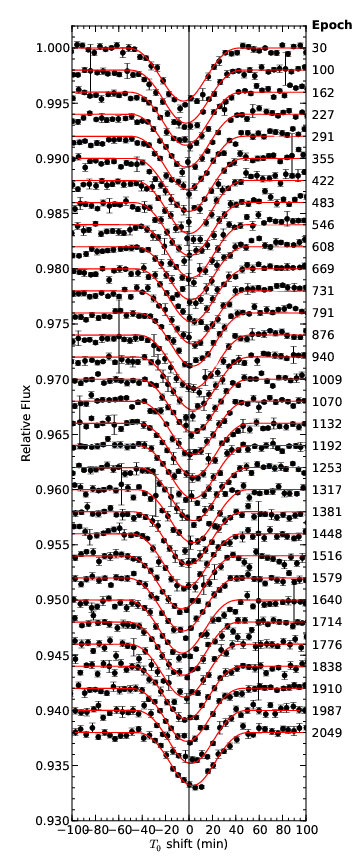}
\caption{The transit for each segment of the KIC9761199 light curve,
with best fit model overplotted. For details see \textit{Fig. 10}.}
\label{Fig. 14}
\end{figure}  


\subsection{Exoplanet candidate KIC5972334}
The Kepler exoplanet candidate KIC5972334 is another interesting 
object in our sample. We first obtained the light curve using 
the period $P=0.7085982$ days reported by \textit{Batalha et al. (2013)}.
It is shown in \textit{Fig. 15}.
We did not find a convincing transit signal at this orbital period.
We searched for periods and obtained
$P=15.35747382$ days using the PDM and $P=15.36051120$ days using
the FA method. We did not confirm the preliminary orbital period. 
According to these results the light curve of KIC5972334 was reanalysed 
with $P=15.35747382$ days.
\textit{Steffen et al. (2010)} suggested that this star shows
transits from two exoplanets with orbital period $P=15.359$ days and 
$P=2.420$ days. We confirmed the first period, but could not confirm 
the second object via PDM/FA period analysis.
The transit light curve folded with the period $P=15.35747382$ days is shown in
\textit{Fig. A20 -- left panel}. 
It is the deepest transit in our sample (0.0135) and has a typical
U-shape. We did not detect significant differences between the odd and 
even transits in this system, nor significant pre-transit brightening and post-transit brightening, nor
significant variability in transit depth. This expected due to the long period nature of the candidate.
Subsequently, we fitted the light curve with a model light curve (Fig. A20 -- see
left panel)
and obtained transit residuals of this
exoplanet candidate (Fig. A20 -- mid panel).
The residuals show that this transit can be reproduced well with the planetary model.
We obtained quadratic limb darkening coefficients of $q_1 = 0.33(8)$ and $q_2 = 0.4(1)$. 
The best fit system parameters are: mid transit time $T_c=454.9191(1)$, period $P=15.35877$ days, 
planet-to-star radius ratio $R_p/R_s=0.118(1)$, normalised semi-major axis $(R_p+R_s)/a=0.038(1)$, and 
line-of-sight inclination $i=89.1(1)$ (deg). The parameter $R_p/R_s=0.01292$, presented by \textit{Batalha et al. (2013)}, is very different from what we got for $R_p/R_s$, because it is a different candidate at a different period.

\begin{figure}[!h]
\centering
\includegraphics[width=80mm]{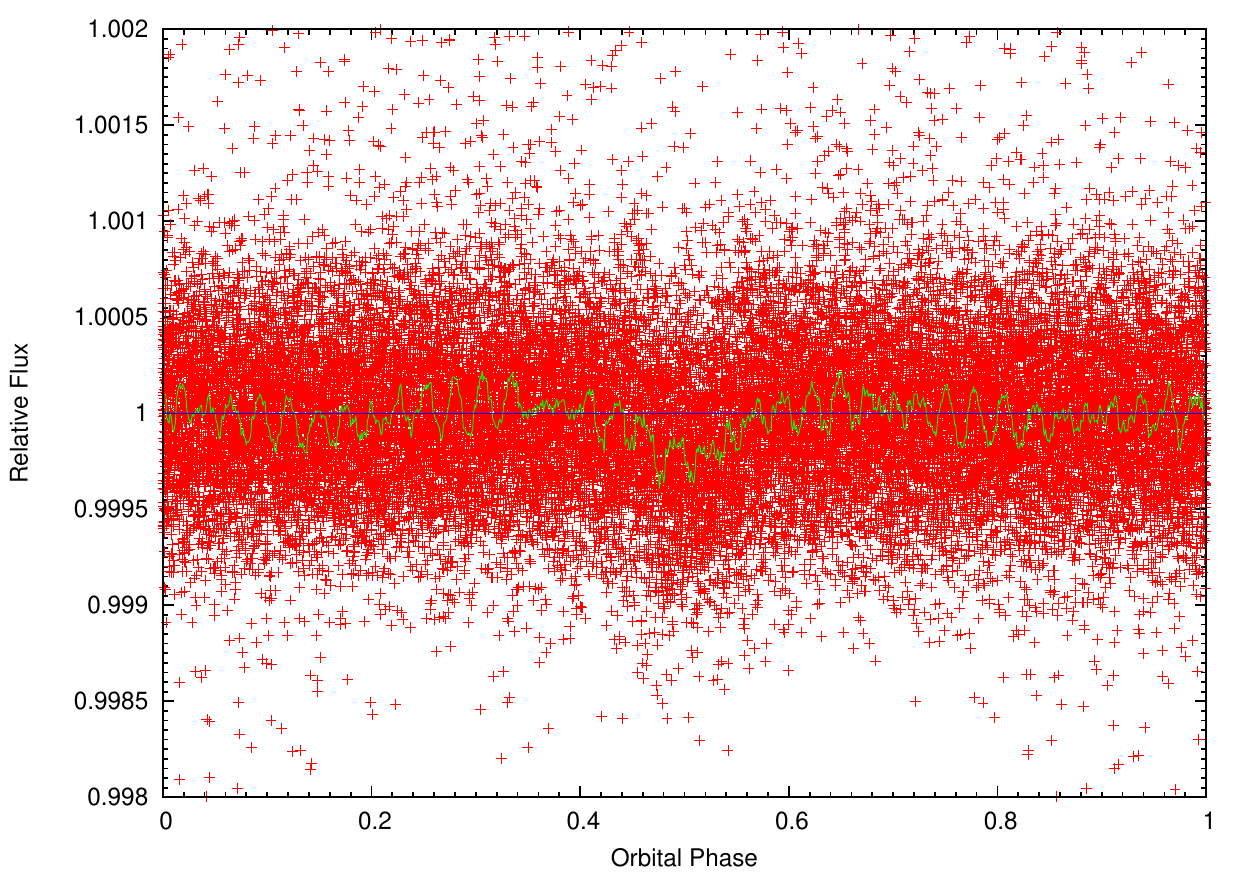}
\caption{The light curve of KIC5972334 phased to the orbital period $P=0.7085982$ days, reported in \textit{Batalha et al.
(2013)}.}
\label{Fig. 9}
\end{figure}

Finaly we searched for possible long-term changes of the orbital 
period. \textit{Fig. A20 -- right panel} shows that there is 
an indication for the long-term period change, with
$\beta=33.3316 \pm 11.9749$ d/Myr, which means that the period is potentially increasing.
Since the PDM analysis indicated a potential period variation, 
we broke the light curve into segments of 2000 points in this case, and re-performed the
$T_c$-free MCMC analysis. However, this analysis did not find any TTV signal.


\subsection{Other exoplanet candidates}     
The candidates KIC6934291, KIC4055304, KIC10024051, KIC8235924, KIC10975146, KIC10028535, KIC10468885, KIC11600889, KIC8278371, KIC5513012, KIC10319385 and KIC9473078 did not show any peculiarities in their transit light curves, nor residuals
(Fig. A6, A7, A8, A9, A11, A12, A13, A14, A15, A16, A17 and A19 -- left and mid panels). 
These transits have typically shallow U-shapes.
We did not detect significant differences between the odd and even 
transits in these systems. We fitted the light curves of these candidates with a 
theoretical light curve. The best fit parameters of all candidates are 
summarized in \textit{Table 2}.
We also improved the orbital periods 
reported in \textit{Batalha et al. (2013)} via PDM and FA methods. 
Finaly we searched for possible long-term changes of the orbital
periods. They are all listed in \textit{Table 3}. \textit{Fig. A6, A7, A8, A9, A11, A12, A13, A14, A15, A16, A17 and A19 -- right
panels} show that there are no significant evidences for the long-term 
period changes. In case of the exoplanet candidate KIC10468885 we detected 
an indication for the long-term period change with
$\beta=3.0$ d/Myr, which means that the period is increasing (Fig. A13 -- 
right panel), however, based on the PDM Monte Carlo test, we could not consider this
result as significant enough ($std= \pm 8.7$ d/Myr).

\begin{table*}[!t]   
\centering
\caption{An overview of the best fit parameters. Table contains the best fit qadratic limb darkening coefficients of the host stars ($q_1$ and $q_2$), and the best fit system parameters: mid transit time $T_c$, period $P$, planet-to-star radius ratio $R_p/R_s$, normalised semi-major axis $(R_p+R_s)/a$, and 
line-of-sight inclination $i$.}
\label{Table 2}
\begin{tabular}{cccccccc}
\hline
\hline
KIC Number & $q_1$ & $q_2$ & $T_c$  & $P$   & $R_p/R_s$ & $(R_p+R_s)/a$ & $i$\\
	   &       &       & (BKJD) & (days) &           &               & (deg)\\  
\hline
\hline
3848972  & 0.9(1)    & 0.9(3)  & 534.34045(5) & 0.37053  & 0.18(8)   & 0.71(4)  & 49(6)\\
8561063  & 0.0001(8) & 0.7(1)  & 488.7930(1)  & 0.45329  & 0.044(4)  & 0.2(1)   & 76(2)\\
6666233  & 0.26(4)   & 0.8(1)  & 546.2643(5)  & 0.51241  & 0.016(2)  & 0.27(1)  & 85(4)\\
6047498  & 0.94(5)   & 0.95(4) & 534.1331(3)  & 0.51873  & 0.3(2)    & 0.76(5)  & 43(5)\\
9030447  & 0.47(2)   & 0.99(2) & 534.478(1)   & 0.56677  & 0.0091(1) & 1.000(8) & 85(1)\\
6934291  & 0.1(1)    & 0.2(2)  & 523.6117(4)  & 0.56786  & 0.018(8)  & 0.3(1)   & 74(14)\\
4055304  & 0.0006(3) & 0.2(2)  & 546.2270(3)  & 0.57104  & 0.015(9)  & 0.24(1)  & 85(5)\\
10024051 & 0.55(2)   & 0.03(3) & 553.9052(3)  & 0.57737  & 0.0187(4) & 0.25(1)  & 86(3)\\
8235924  & 0.02(2)   & 0.3(3)  & 459.5745(4)  & 0.58800  & 0.018(3)  & 0.70(1)  & 47(1)\\
11774303 & 0.118(9)  & 0.7(2)  & 457.3181(7)  & 0.61408  & 0.07(4)   & 0.8(1)   & 37(11)\\
10975146 & 0.4(1)    & 0.8(1)  & 496.9595(4)  & 0.63133  & 0.0184(7) & 0.24(1)  & 87(3)\\
10028535 & 0.1(1)    & 0.2(2)  & 454.9020(7)  & 0.66309  & 0.017(1)  & 0.21(1)  & 86(3)\\
10468885 & 0.03(3)   & 0.5(5   & 545.814(1)   & 0.66407  & 0.015(1)  & 0.26(6)  & 79(10)\\
11600889 & 0.29(7)   & 0.96(3) & 550.4423(4)  & 0.66931  & 0.011(1)  & 0.40(9)  & 74(12)\\
8278371  & 0.06(5)   & 0.88(9) & 484.1484(8)  & 0.67737  & 0.0086(5) & 0.42(9)  & 73(15)\\
5513012  & 0.28(4)   & 0.87(3) & 364.5110(5)  & 0.67934  & 0.0146(9) & 0.25(2)  & 84(5)\\
10319385 & 0.02(1)   & 0.7(1)  & 584.0281(2)  & 0.68921  & 0.014(1)  & 0.79(7)  & 40(1)\\
9761199  & 0.94(5)   & 0.96(1) & 533.1669(1)  & 0.69202  & 0.22(9)   & 0.43(3)  & 68(2)\\
9473078  & 0.7(4)    & 0.3(3)  & 574.852(1)   & 0.69384  & 0.010(4)  & 0.83(5)  & 35(5)\\
5972334  & 0.33(8)   & 0.4(1)  & 454.9191(1)  & 15.35877 & 0.118(1)  & 0.038(1) & 89.1(1)\\
\hline
\end{tabular}
\end{table*}


\section{Summary and conclusions} 
Our main aim was to search for comet-like tails similar to 
KIC012557548b and for long-term orbital period variations. 
We were curious about frequency of comet-like tail formation among short-period 
Kepler exoplanet candidates. Close-in exoplanets, like KIC012557548b, are subjected to the greatest planet-star interactions.
We therefore concentrated on the short-period exoplanet
candidates with periods similar to KIC012557548b.
We chose 20 exoplanet candidates observed 
by the Kepler mission, with the shortest orbital periods,
ranging from 0.370 up to 0.708 days. 
We first improved the preliminary orbital periods (Batalha et al. 2013) and obtained the transit light curves of our exoplanet candidates.
Subsequently, we searched for the signatures of a circum-planetary material
in these light curves.
For this purpose the final transit light curve of each planet was
fitted with a theoretical light curve, and the residuals were examined for abnormalities.
We also searched for possible long-term changes of the orbital period using the method of phase dispersion minimization (PDM).
For candidates where the PDM analysis indicated potential period variations,
we broke the light curve into segments of 2000/4000 points, and re-performed the
$T_c$-free MCMC analysis. To test the effectiveness of our techniques, we first applied
this procedure to KIC012557548b. The light curve of this exoplanet is
peculiar, caused by a pre-transit brightening, post-transit brightening and strong variability in the transit depth.
Subsequently, we fitted the observed light curve with the theoretical model light curve,
assuming a spherical planet, and obtained transit
residuals. We observed a significant fluctuation in these residuals. 
We then searched for a long-term orbital period variation using the PDM method. Previously, \textit{Budaj (2013)} obtained $\beta=0.3\pm0.5$ d/Myr using 14 quarters of Kepler data, indicating no change in the transit period. We repeated the analysis using 17 quarter of data, revealing no significant long-term period variation.

We then analyzed the transit light curves and residuals of our sample similarly as per KIC012557548b. We searched for significant fluctuations in these residuals, strong variabilities in the transit depth, pre-transit and post-transit brightenings in light curves, which could indicate presence of the circum-planetary material. In 8 cases out of 20 we found some interesting peculiarities (KIC3848972, KIC8561063, KIC6666233, KIC6047498, KIC9030447, KIC11774303, KIC9761199, KIC5972334). We found 4 cases of exoplanet candidates with relatively deep V-shape of the transits (KIC3848972, KIC8561063, KIC6047498, KIC9761199). We detected differences between the odd and even transits in 5 cases (KIC3848972, KIC6666233, KIC6047498, KIC11774303, KIC9761199). In cases of KIC3848972, KIC8561063, KIC6666233, KIC6047498, KIC11774303 and KIC9761199 we consider the eclipsing binary
alternative as possible. Excepting KIC11774303, the PDM/FA based two times alias orbital period is also possible. We found 2 cases of exoplanet candidates with sinusoidal out-of-transit periodic background variability, which is very similar to an RS Canum Venaticorum-like distortion wave (KIC3848972 and KIC9761199). In addition, the sinusoidal background variability shifted in phase during the Kepler observations, suggesting that it is not due to Doppler beaming. The background variations may potentially be due to stellar spot modulation. The exoplanet candidate KIC9030447 is also peculiar, because its light curve shows a small brightening, approximately half a phase from the transit, and a secondary eclipse-like signal approximately at phase 0.85. Further we discovered that these additional features of the light curve slighly shifted in phase during the Kepler observations. Moreover, the secondary eclipse-like signal is changing its depth. That is why we consider the eclipsing binary on eccentric orbit alternative as more possible as exoplanet transits or stellar pulsations.  

\begin{table*}[!t]   
\centering
\caption{An overview of improved orbital periods. Exoplanet candidates signed
at their  KIC Number with
$\ast$ may have twice as long orbital period. 
We did not confirm the preliminary orbital period in case of candidate KIC5972334. Table also contains the $\beta$ parameter.}
\label{Table 2}
\begin{tabular}{cccccc}\hline
\hline
KIC Number & Period by \textit{Batalha et al. (2013)} & Period by PDM & Period by FA & Period by MCMC & $\beta$\\
& (days) & (days) & (days) & (days) & (d/Myr)\\
\hline
\hline
3848972$^{\ast}$ &  0.3705290  &  0.37052870  & 0.37052750  & 0.37053 & -0.8  $\pm$ 1.4\\              
8561063$^{\ast}$ &  0.4532875  &  0.45328705  & 0.45328830  & 0.45329 &  0.92 $\pm$ 0.93\\
6666233$^{\ast}$ &  0.5124040  &  0.51240883  & 0.51240860  & 0.51241 & -1.0  $\pm$ 2.3\\
6047498$^{\ast}$ &  0.5187486  &  0.51872845  & 0.51872930  & 0.51873 & -15.0 $\pm$ 1.9\\
9030447  &           0.5667281  &  0.56670873  & 0.56680630 & 0.56677 &  0.9  $\pm$ 8.6\\
6934291  &           0.5678562  &  0.56785838  & 0.56785920 & 0.56786 & -1.0  $\pm$ 4.0\\
4055304  &           0.5710391  &  0.57103788  & 0.57104400 & 0.57104 &  1.0  $\pm$ 49.7\\
10024051 &           0.5773752  &  0.57737694  & 0.57736750 & 0.57737 &  0.9  $\pm$ 1.6\\
8235924  &           0.5879933  &  0.58800171  & 0.58800290 & 0.58800 &  1.0  $\pm$ 1.8\\
11774303 &          0.6140747 &   0.61408403  & 0.61409220  & 0.61408 &  0.9  $\pm$ 2.9\\
10975146 &          0.6313298 &   0.63133287  & 0.63133840  & 0.63133 & -1.0  $\pm$ 1.4\\
10028535 &          0.6630983 &   0.66308772  & 0.66307940  & 0.66309 & -1.0  $\pm$ 3.6\\
10468885 &          0.6640818 &   0.66407462  & 0.66406930  & 0.66407 &  3.0  $\pm$ 8.7\\
11600889 &          0.6693163 &   0.66931924  & 0.66930920  & 0.66931 &  1.0  $\pm$ 1.2\\
8278371  &          0.6773911 &   0.67737580  & 0.67737370  & 0.67737 & -1.0  $\pm$ 3.3\\
5513012  &          0.6793614 &   0.67933835  & 0.67933710  & 0.67934 &  0.9  $\pm$ 3.1\\
10319385 &          0.6892040 &   0.68920953  & 0.68920870  & 0.68921 & -1.0  $\pm$ 10.9\\
9761199$^{\ast}$ & 0.6920069 &   0.69201741  & 0.69204530   & 0.69202 & -3.0  $\pm$ 0.4\\
9473078  &          0.6938521 &   0.69384010  & 0.69383390  & 0.69384 &  1.0  $\pm$ 7.3\\
5972334  &          0.7085982 &  15.35747382 & 15.36051120  & 15.35877 & 33.3 $\pm$ 11.9\\
\hline
\end{tabular}
\end{table*} 

None of the exoplanet candidates shows signs of a comet-like tail. We did not find any significant fluctuations in residuals, strong variabilities in the transit depth, nor pre-transit and post-transit brightenings in light curves. We found only a linear downward/upward residual trend from sine-like background variability in cases of exoplanet candidates KIC3848972 and KIC9761199, and a small residual fluctuation in case of exoplanet candidate KIC9030447 due to the poor fit. It seems that the frequency of comet-like tail formation among short-period Kepler exoplanet candidates is very low. We searched for comet-like tails based on the period criterion. Based on our results we can conclude that the short-period criterion is not enough to cause comet-like tail formation. This result is in agreement with the theory of the thermal wind and planet evaporation (Perez-Becker \& Chiang 2013). This theory predicts that catastrophic evaporation, which leads into creation of a comet-like tail, can occure only in relatively low-mass planets with masses less than that of Mercury. Since all the exoplanet candidates out of 20 have planet-to-star radius ratio higher than 0.003 (see parameter $R_p/R_s$ in Table 2; the value 0.003 approximately valids for the planet Mercury), they are likely to be more massive than Mercury and, consequently, lack the catastrophic evaporation contrary to KIC012557548b. 

We also found 3 cases of candidates (KIC6047498, KIC9761199 and KIC5972334) which showed some changes of the orbital period. In one case (KIC5972334) we observed an orbital period increasing, other exoplanet candidates showed orbital period shortenings. Since the PDM analysis indicated a potential period variation in these cases, we broke the light curve into segments of 2000/4000 points, and re-performed the $T_c$-free MCMC analysis. In case of exoplanet candidate KIC6047498 we found a semi-periodic TTV signal associated with the transit, indicative of an outer companion with an orbital period of $ > 2000$ days. In case of exoplanet candidate KIC9761199 this analysis also confirmed our results from the PDM analysis about long-term period change and revealed a periodic TTV signal with a period of $\sim 1500$ days, indicative of a possible massive outer companion. In case of exoplanet candidate KIC5972334 this analysis did not find any TTV signal. Based on our results we can see that orbital period changes are not caused by comet-like tail disintegration processes, but rather by possible massive outer companions. Moreover, nor KIC012557548b with comet-like tail disintegration, did not show any long-term orbital period variation.
 
We also improved the preliminary orbital periods using PDM/FA methods. We did not confirm the preliminary orbital period in case of exoplanet candidate KIC5972334.
We obtained period $P = 15.35747382$ days using the PDM and $P = 15.36051120$ days using the FA method, which is in the agreement with period $P=15.359$ days found by \textit{Steffen et al. (2010)}. We did not confirm the second transiting object with \textit{P} over 2.420 days in this system, suggested by the same authors. The exoplanet candidates KIC3848972, KIC8561063, KIC6666233, KIC6047498 and KIC9761199 may have twice as long orbital
period. 

\vspace{0.4cm}
\acknowledgements
The authors thank Dr. L. Hamb\'{a}lek and Dr. T. Krej\v{c}ov\'{a} for 
the technical assistance, comments and discussions.
This work was supported by the VEGA grants of the Slovak Academy of
Sciences Nos. 2/0143/14, 2/0038/13, by the Slovak Research
and Development Agency under the contract No. APVV-0158-11, and 
by the realization of the
Project ITMS No. 26220120029, based on the Supporting Operational
Research and Development Program financed from the European Regional
Development Fund. JB was supported by the Australian Research Council through DP120101792.
GZ thanks Chelsea X. Huang for her contribution to the light curve fitting program.   

\clearpage
\newpage
\appendix
\onecolumn
\section{An overview of light curves, transit residuals, and long-term 
changes of the orbital periods}

\begin{figure*}[!h]
\centering
\centerline{
\includegraphics[width=58mm]{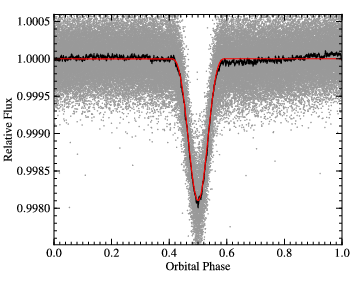}
\includegraphics[width=58mm]{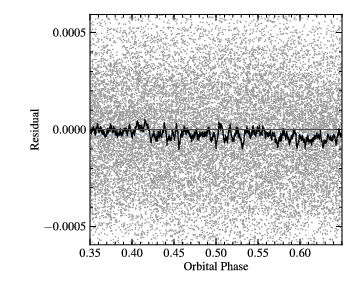}
\includegraphics[width=65mm]{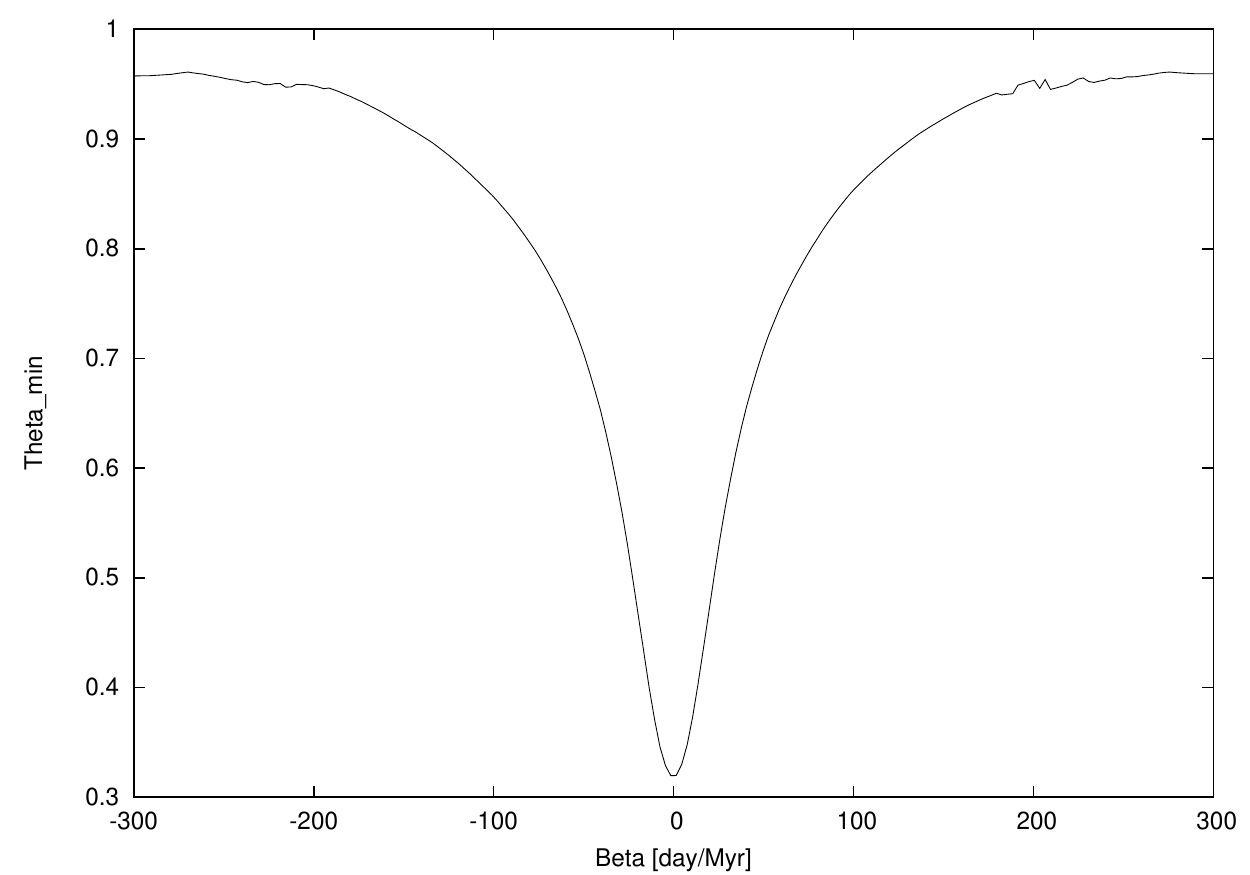}}
\caption{The light curve (left panel), transit residuals
(mid panel), and the long-term change of orbital period (right
panel) of the exoplanet candidate KIC3848972.}
\label{Fig. A1}
\end{figure*}

\begin{figure*}[!h]
\centering
\centerline{
\includegraphics[width=58mm]{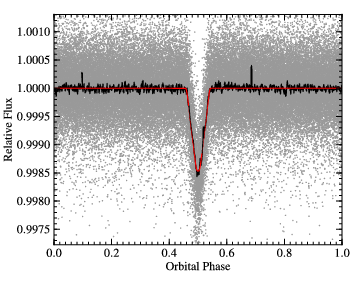}
\includegraphics[width=58mm]{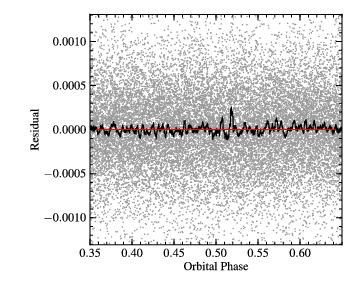}
\includegraphics[width=65mm]{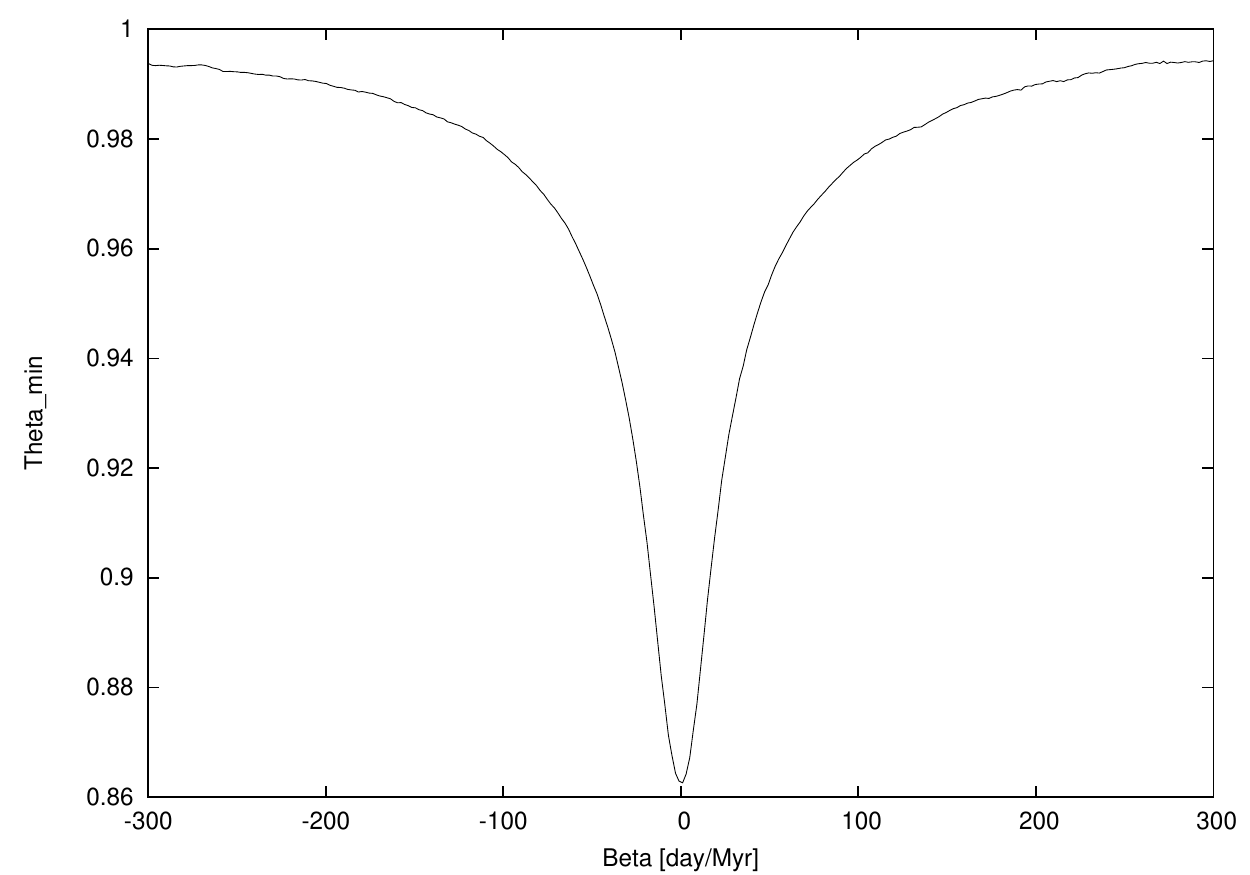}} 
\caption{The light curve (left panel), transit residuals
(mid panel), and the long-term change of orbital period (right
panel) of the exoplanet candidate KIC8561063.}
\label{Fig. A2}
\end{figure*}

\begin{figure*}[!h]
\centering
\centerline{
\includegraphics[width=58mm]{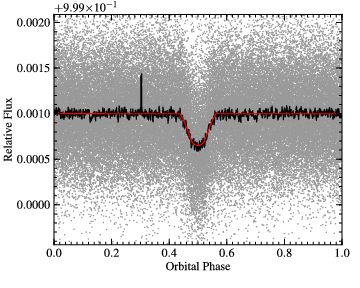}
\includegraphics[width=58mm]{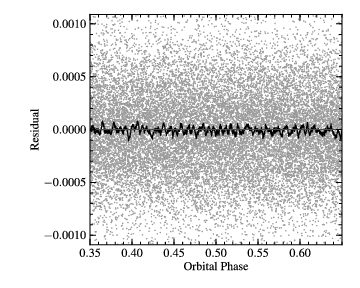}  
\includegraphics[width=65mm]{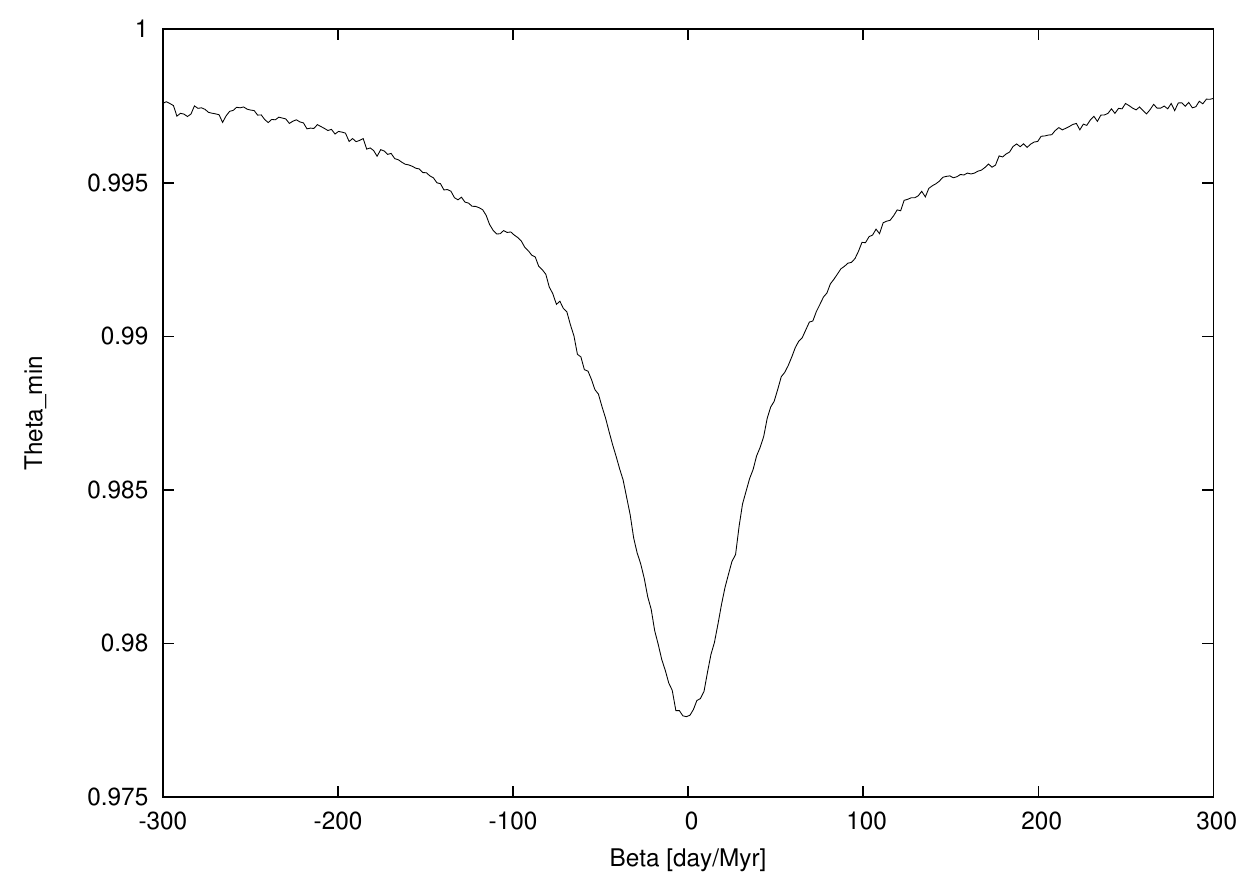}}
\caption{The light curve (left panel), transit residuals
(mid panel), and the long-term change of orbital period (right
panel) of the exoplanet candidate KIC6666233.}
\label{Fig. A3}
\end{figure*}

\begin{figure*}[!h]
\centering
\centerline{
\includegraphics[width=58mm]{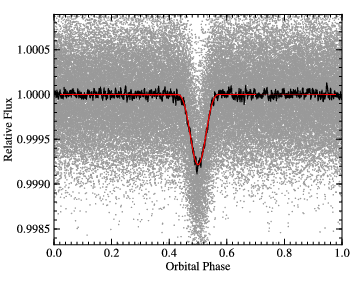}
\includegraphics[width=58mm]{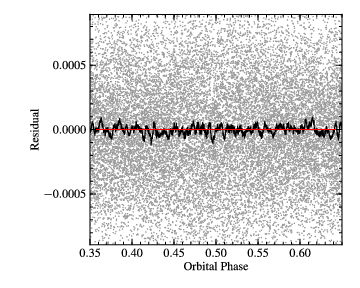}
\includegraphics[width=65mm]{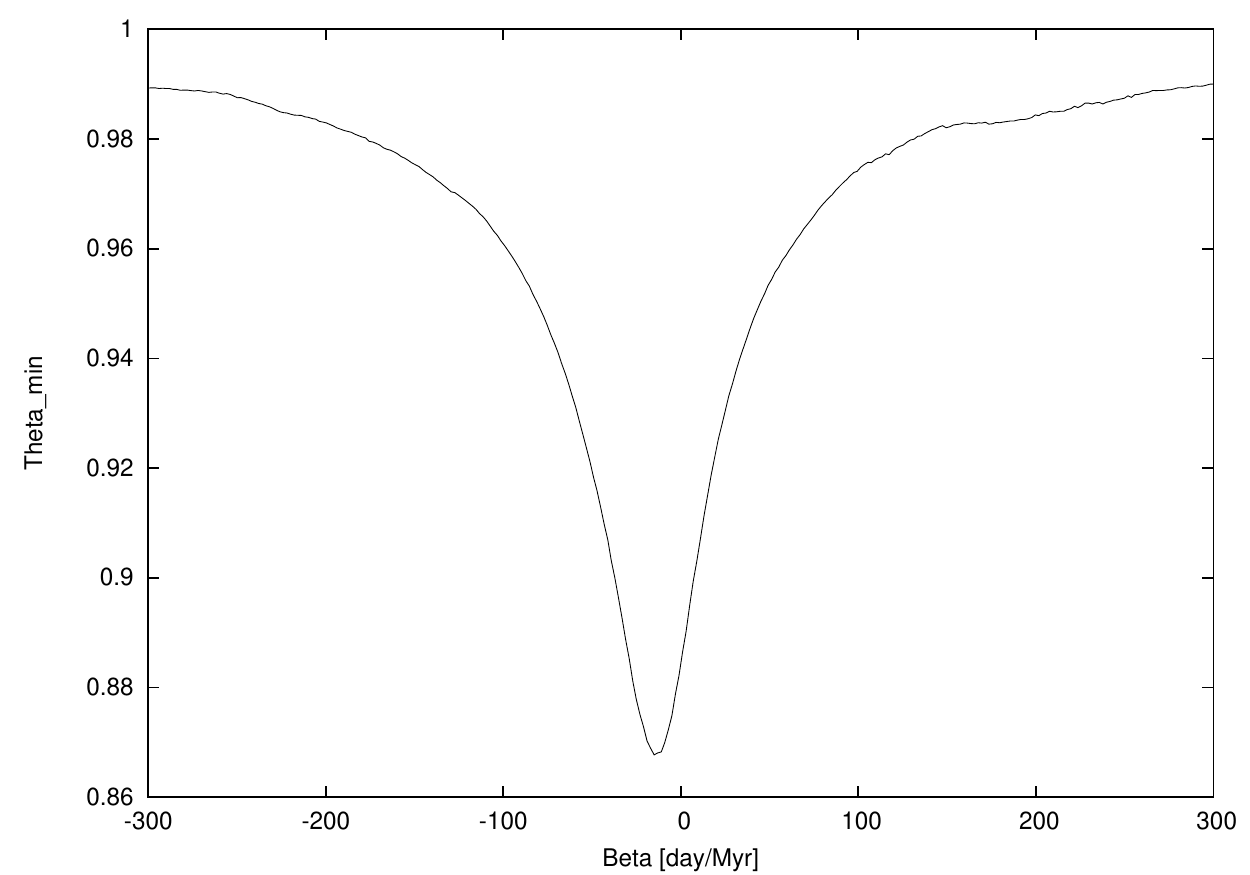}}
\caption{The light curve (left panel), transit residuals
(mid panel), and the long-term change of orbital period (right
panel) of the exoplanet candidate KIC6047498.}
\label{Fig. A4}
\end{figure*}

\begin{figure*}[!h]
\centering
\centerline{
\includegraphics[width=58mm]{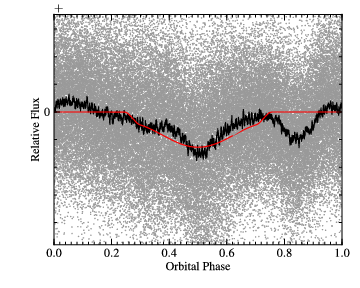}
\includegraphics[width=58mm]{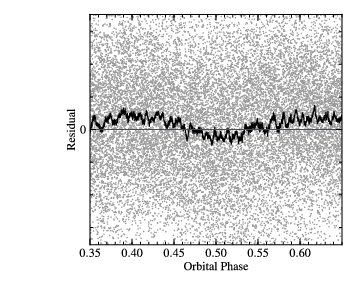}
\includegraphics[width=65mm]{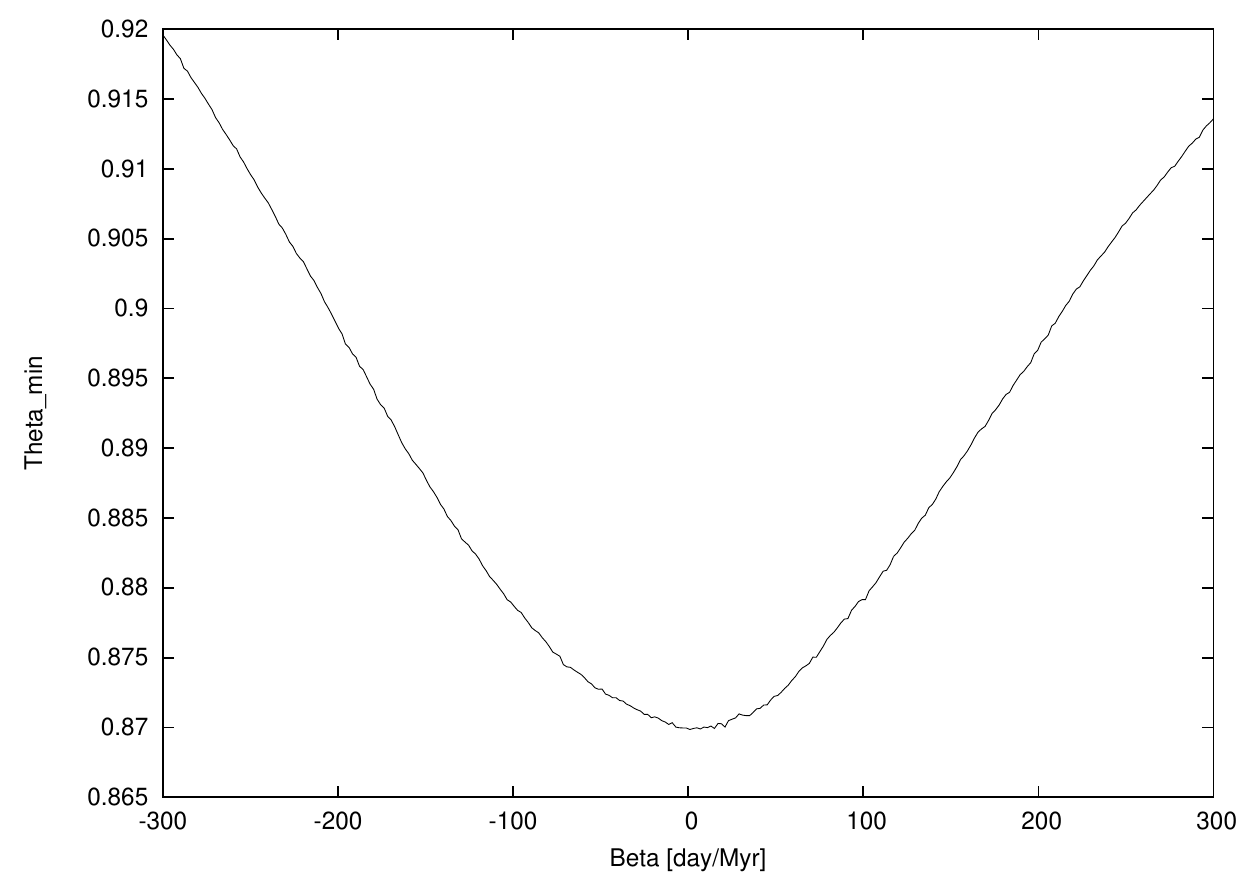}}
\caption{The light curve (left panel), transit residuals
(mid panel), and the long-term change of orbital period (right
panel) of the exoplanet candidate KIC9030447.}
\label{Fig. A5}
\end{figure*}

\begin{figure*}[!h] 
\centering
\centerline{
\includegraphics[width=58mm]{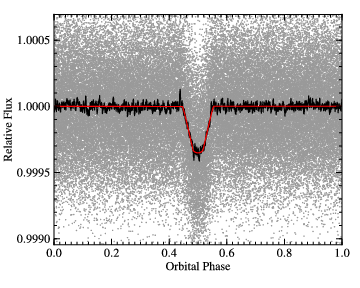}
\includegraphics[width=58mm]{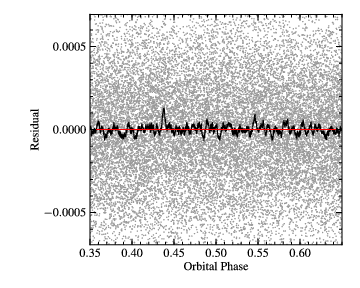}
\includegraphics[width=65mm]{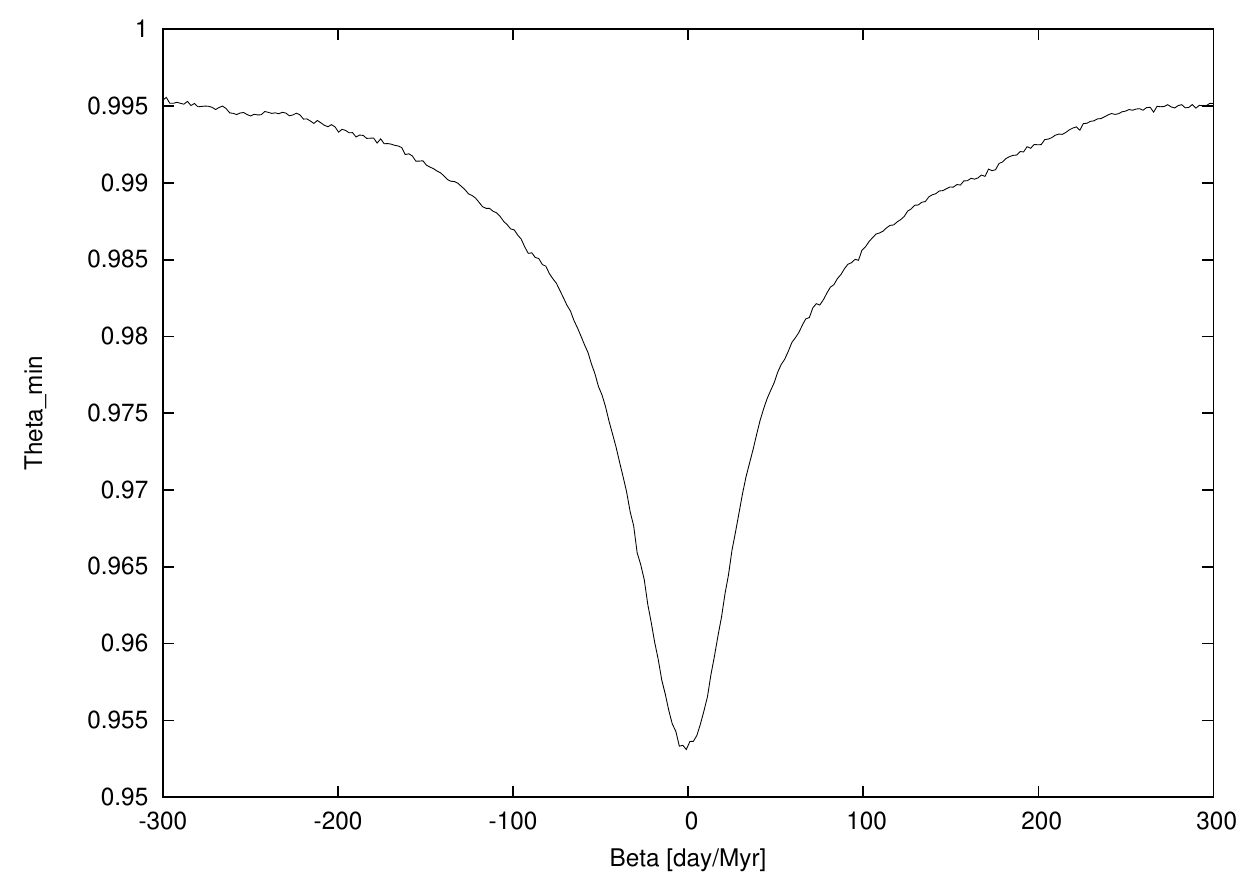}}
\caption{The light curve (left panel), transit residuals
(mid panel), and the long-term change of orbital period (right
panel) of the exoplanet candidate KIC6934291.}
\label{Fig. A6}
\end{figure*}

\begin{figure*}[!h] 
\centering
\centerline{
\includegraphics[width=58mm]{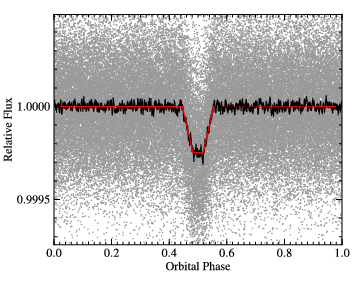}
\includegraphics[width=58mm]{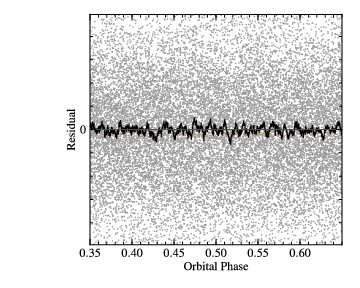}
\includegraphics[width=65mm]{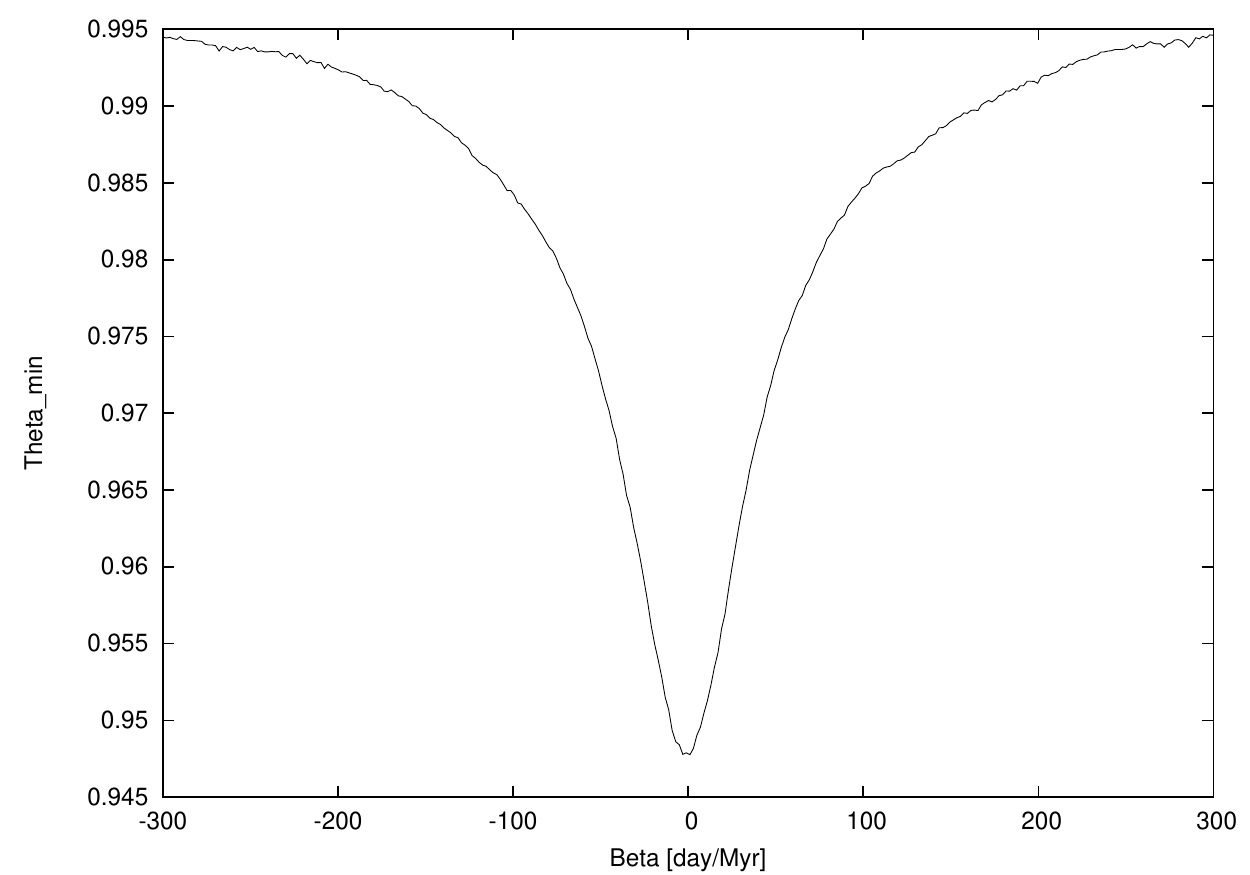}}
\caption{The light curve (left panel), transit residuals
(mid panel), and the long-term change of orbital period (right
panel) of the exoplanet candidate KIC4055304.}
\label{Fig. A7}
\end{figure*}

\begin{figure*}[!h] 
\centering
\centerline{
\includegraphics[width=58mm]{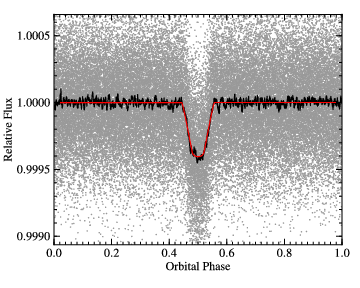}
\includegraphics[width=58mm]{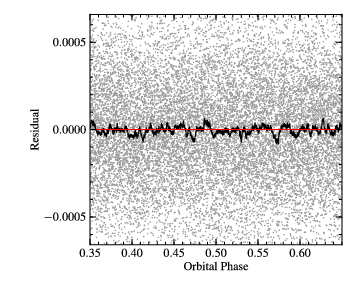}
\includegraphics[width=65mm]{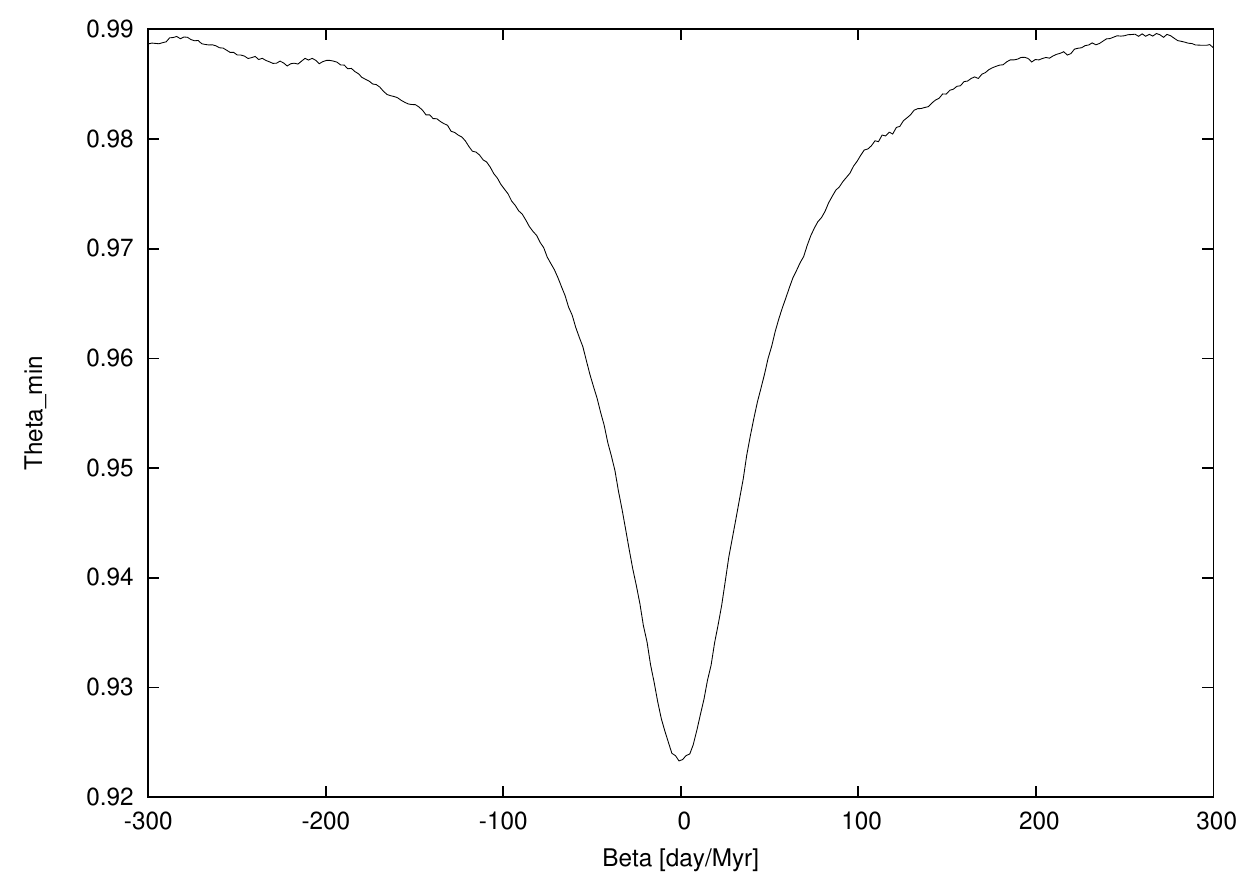}}
\caption{The light curve (left panel), transit residuals
(mid panel), and the long-term change of orbital period (right
panel) of the exoplanet candidate KIC10024051.}
\label{Fig. A8}
\end{figure*}

\begin{figure*}[!h] 
\centering
\centerline{
\includegraphics[width=58mm]{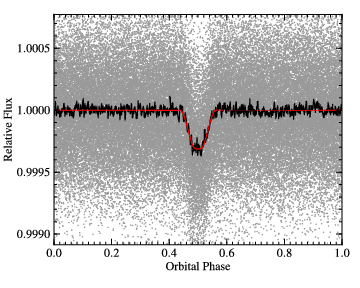}
\includegraphics[width=58mm]{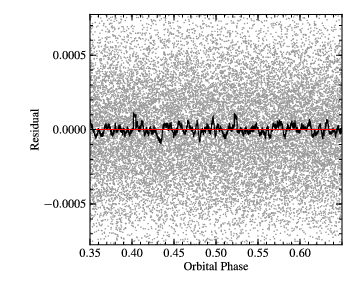}
\includegraphics[width=65mm]{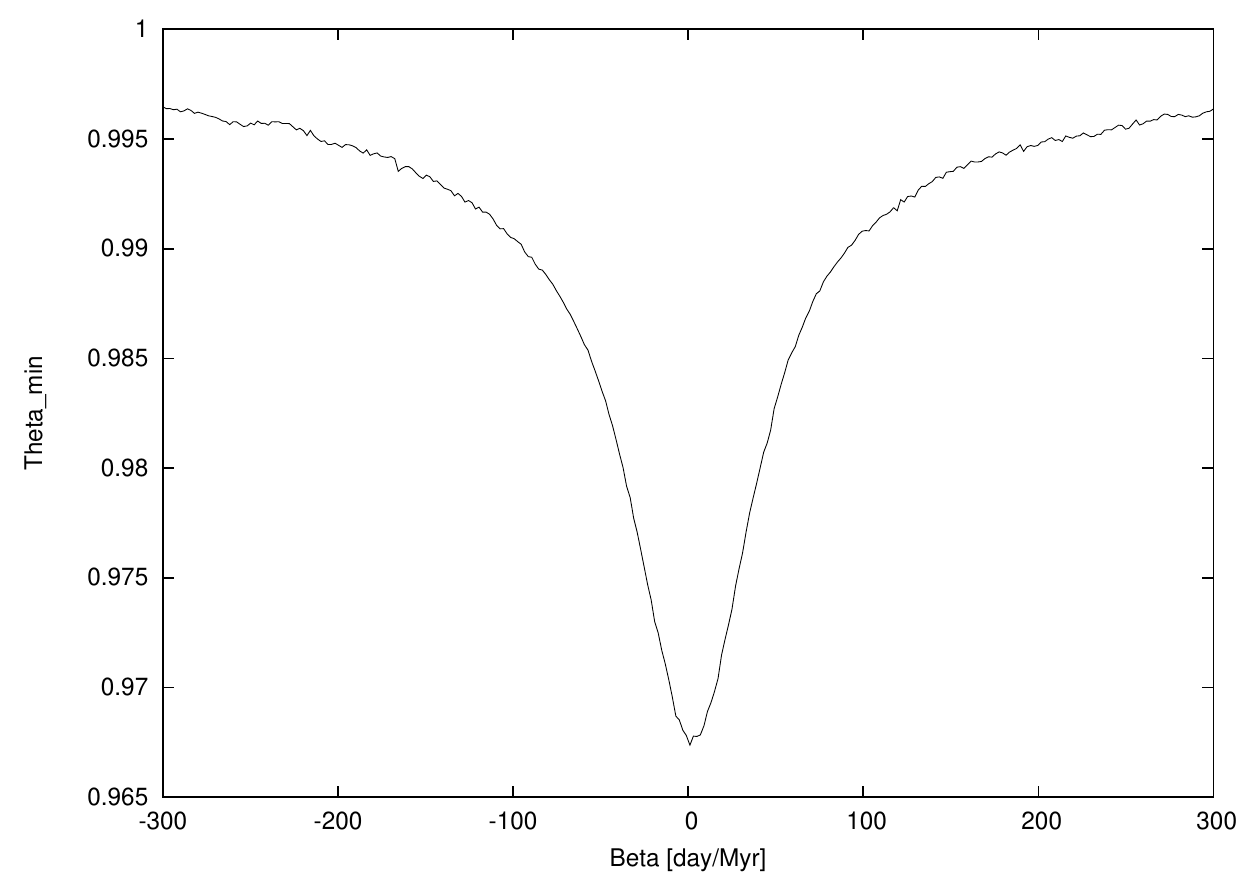}}
\caption{The light curve (left panel), transit residuals
(mid panel), and the long-term change of orbital period (right
panel) of the exoplanet candidate KIC8235924.}
\label{Fig. A9}
\end{figure*}

\begin{figure*}[!h] 
\centering
\centerline{
\includegraphics[width=58mm]{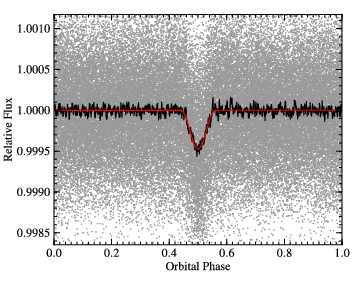}
\includegraphics[width=58mm]{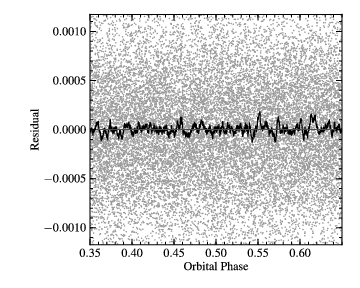}
\includegraphics[width=65mm]{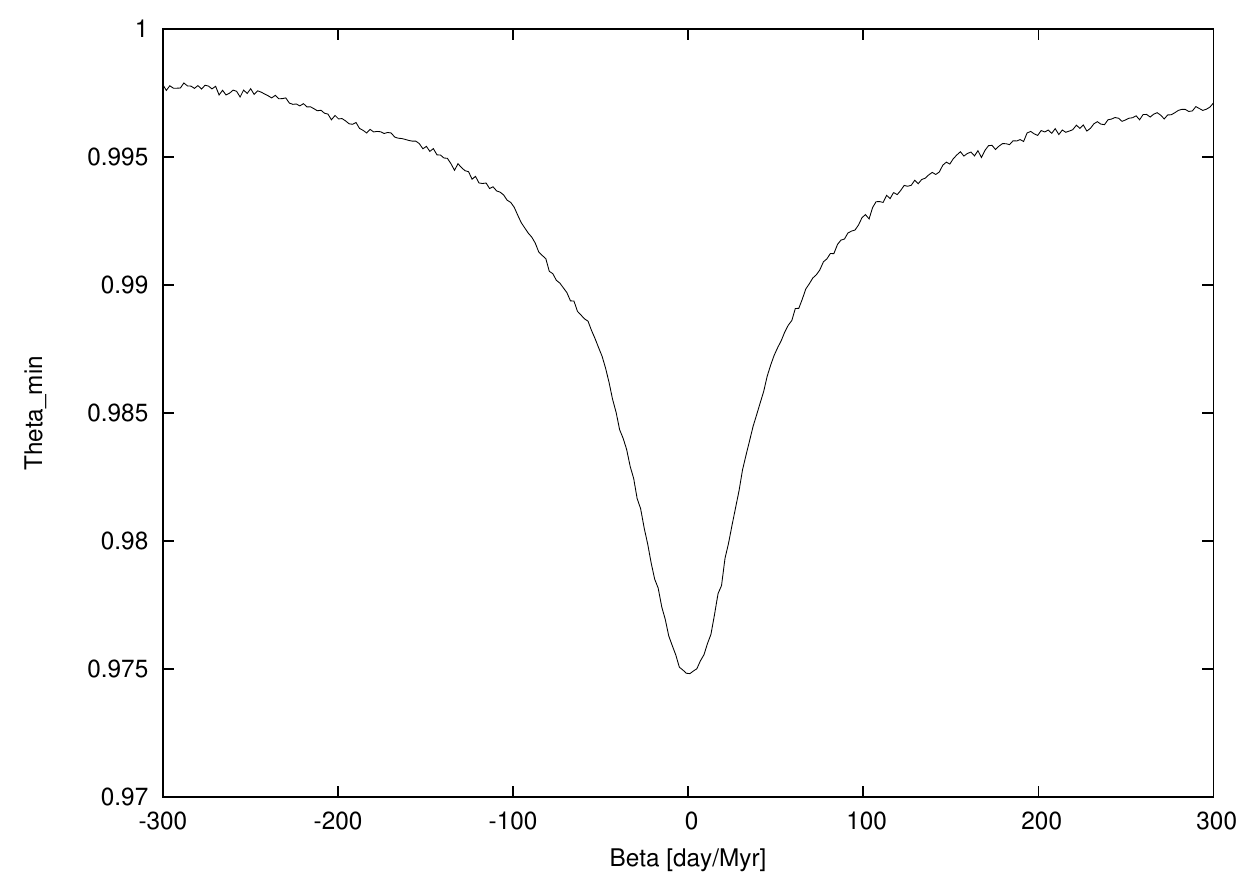}}
\caption{The light curve (left panel), transit residuals
(mid panel), and the long-term change of orbital period (right
panel) of the exoplanet candidate KIC11774303.}
\label{Fig. A10}
\end{figure*}

\begin{figure*}[!h]
\centering
\centerline{
\includegraphics[width=58mm]{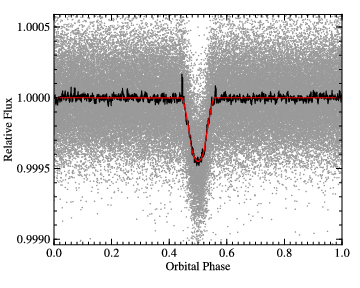}
\includegraphics[width=58mm]{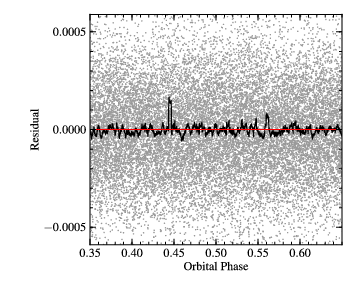}
\includegraphics[width=65mm]{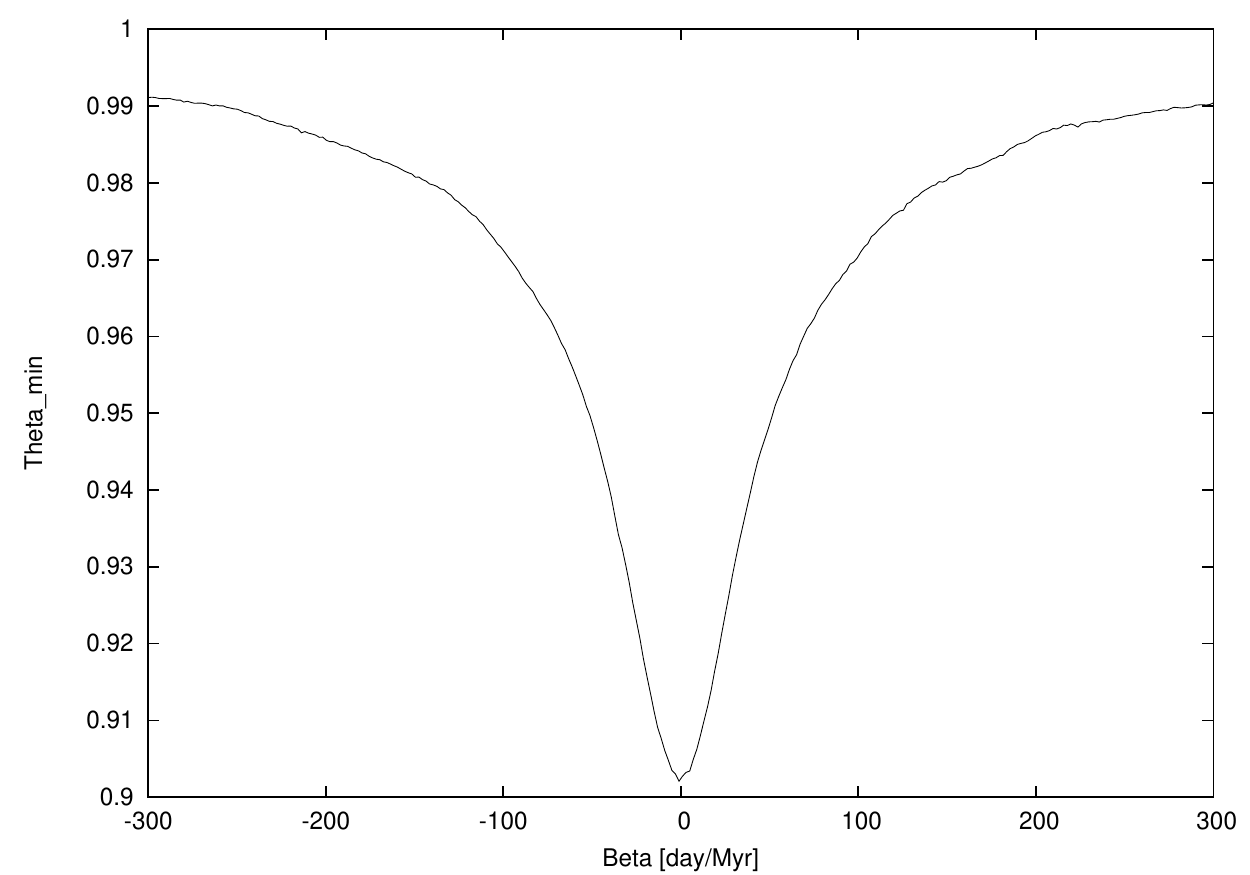}}
\caption{The light curve (left panel), transit residuals
(mid panel), and the long-term change of orbital period (right
panel) of the exoplanet candidate KIC10975146.}
\label{Fig. A11}
\end{figure*}

\begin{figure*}[!h]
\centering
\centerline{
\includegraphics[width=58mm]{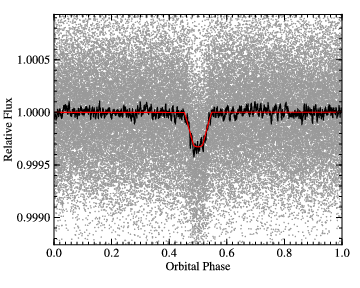}
\includegraphics[width=58mm]{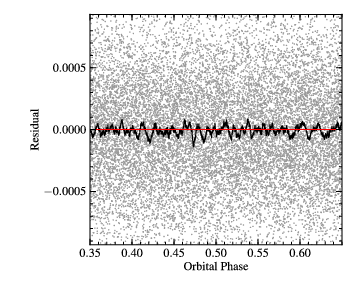} 
\includegraphics[width=65mm]{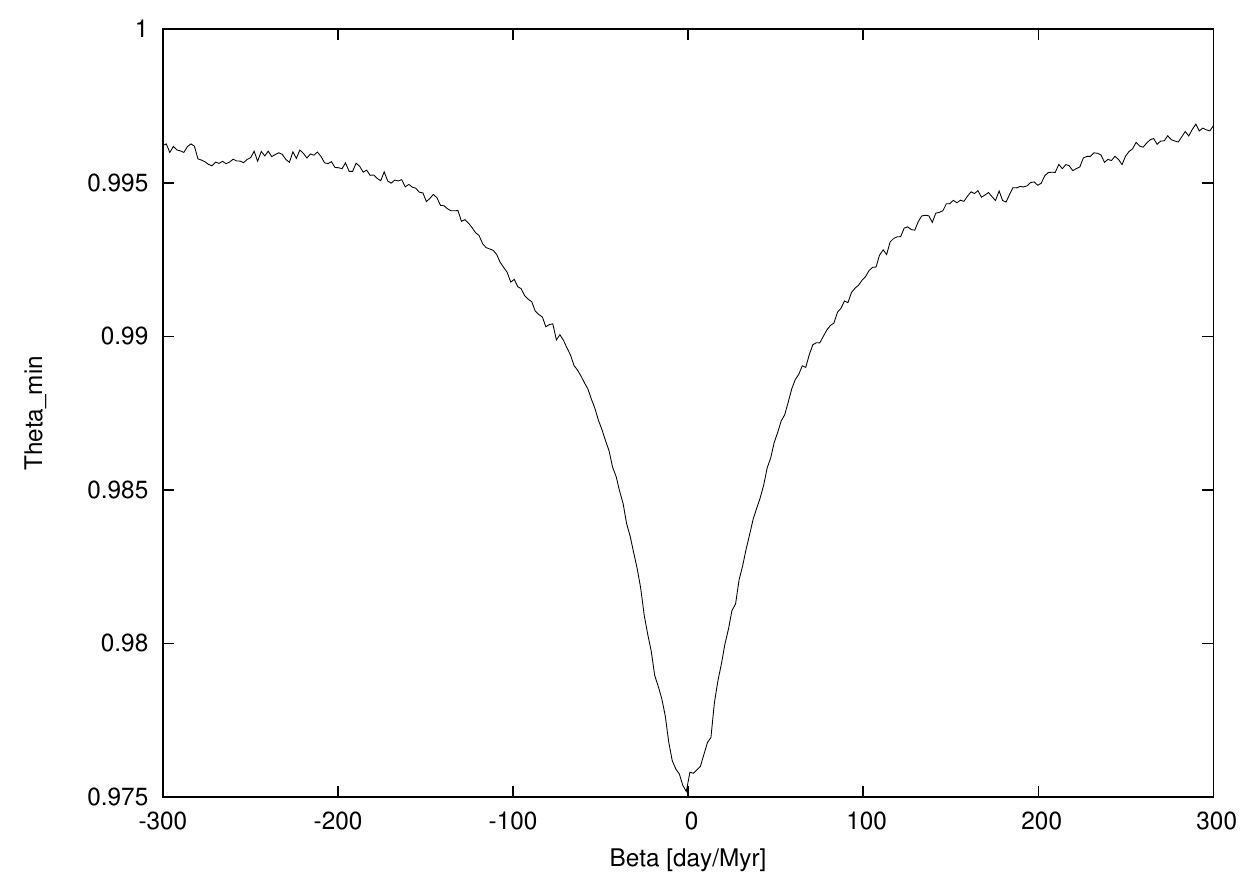}}
\caption{The light curve (left panel), transit residuals
(mid panel), and the long-term change of orbital period (right
panel) of the exoplanet candidate KIC10028535.}
\label{Fig. A12}
\end{figure*}

\newpage

\begin{figure*}[!h]
\centering
\centerline{
\includegraphics[width=58mm]{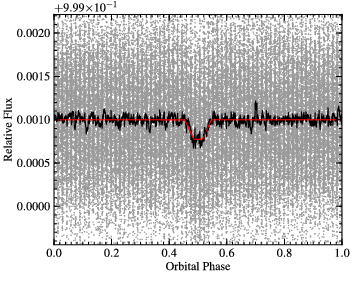}
\includegraphics[width=58mm]{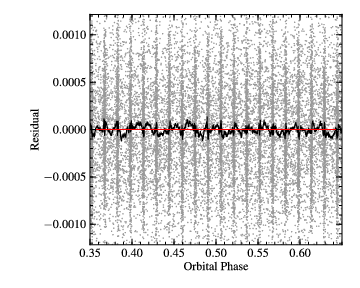}
\includegraphics[width=65mm]{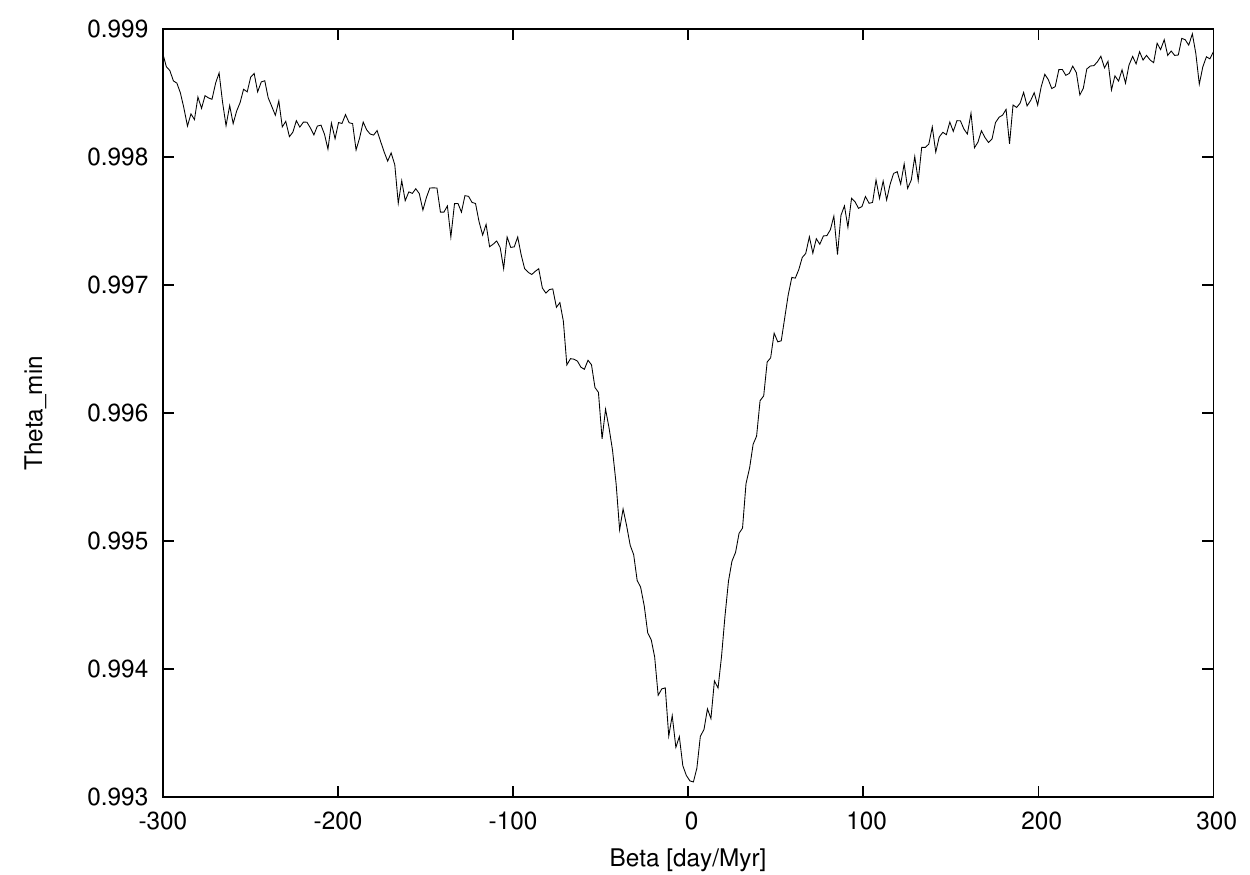}}
\caption{The light curve (left panel), transit residuals
(mid panel), and the long-term change of orbital period (right
panel) of the exoplanet candidate KIC10468885.}
\label{Fig. A13}
\end{figure*}

\begin{figure*}[!h]
\centering
\centerline{
\includegraphics[width=58mm]{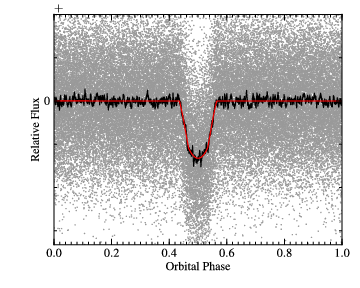}
\includegraphics[width=58mm]{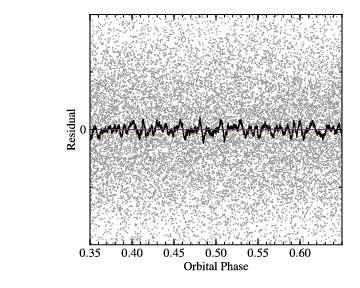}
\includegraphics[width=65mm]{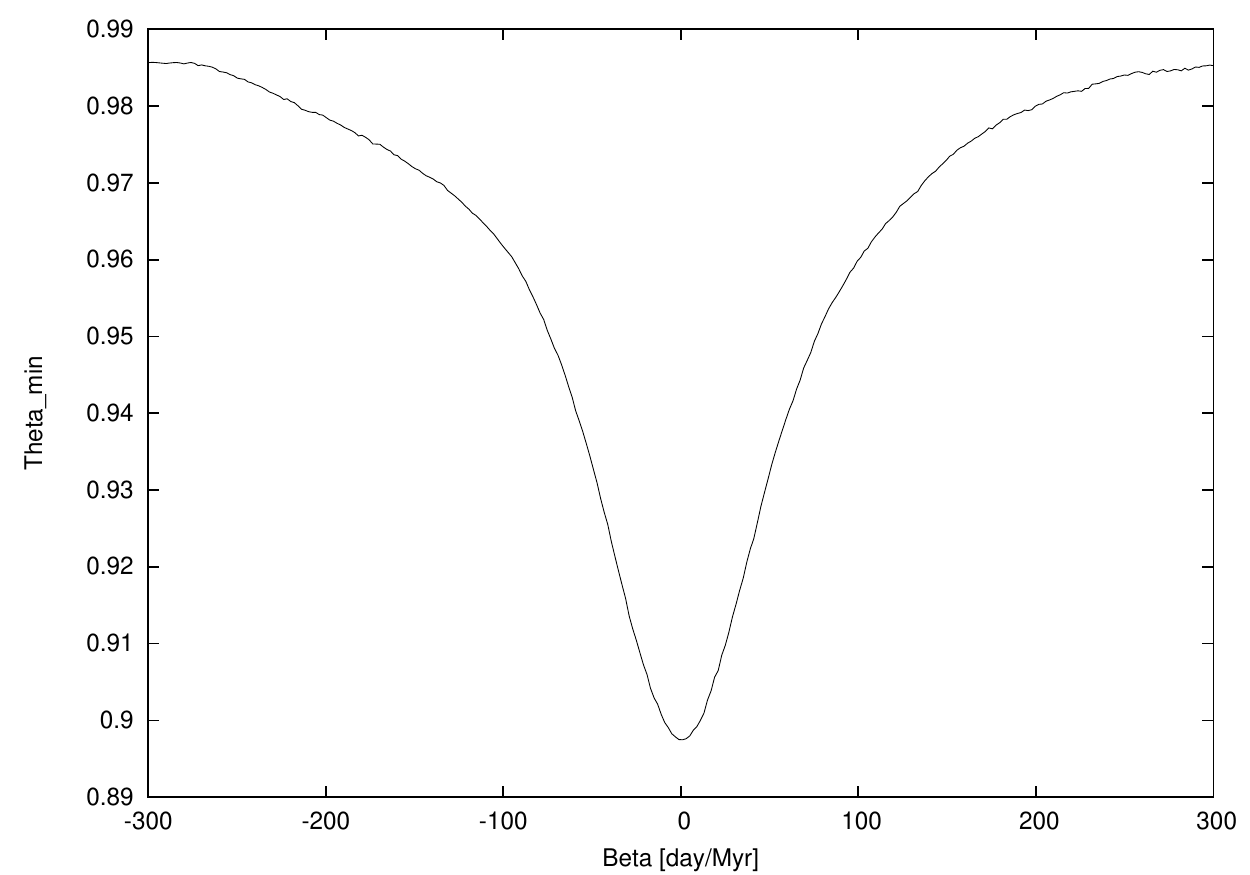}}
\caption{The light curve (left panel), transit residuals
(mid panel), and the long-term change of orbital period (right
panel) of the exoplanet candidate KIC11600889.}
\label{Fig. A14}
\end{figure*}

\begin{figure*}[!h]
\centering
\centerline{
\includegraphics[width=58mm]{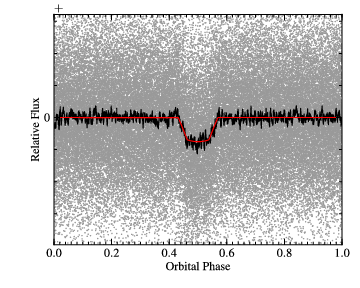}
\includegraphics[width=58mm]{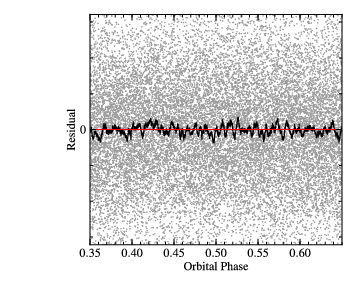}
\includegraphics[width=65mm]{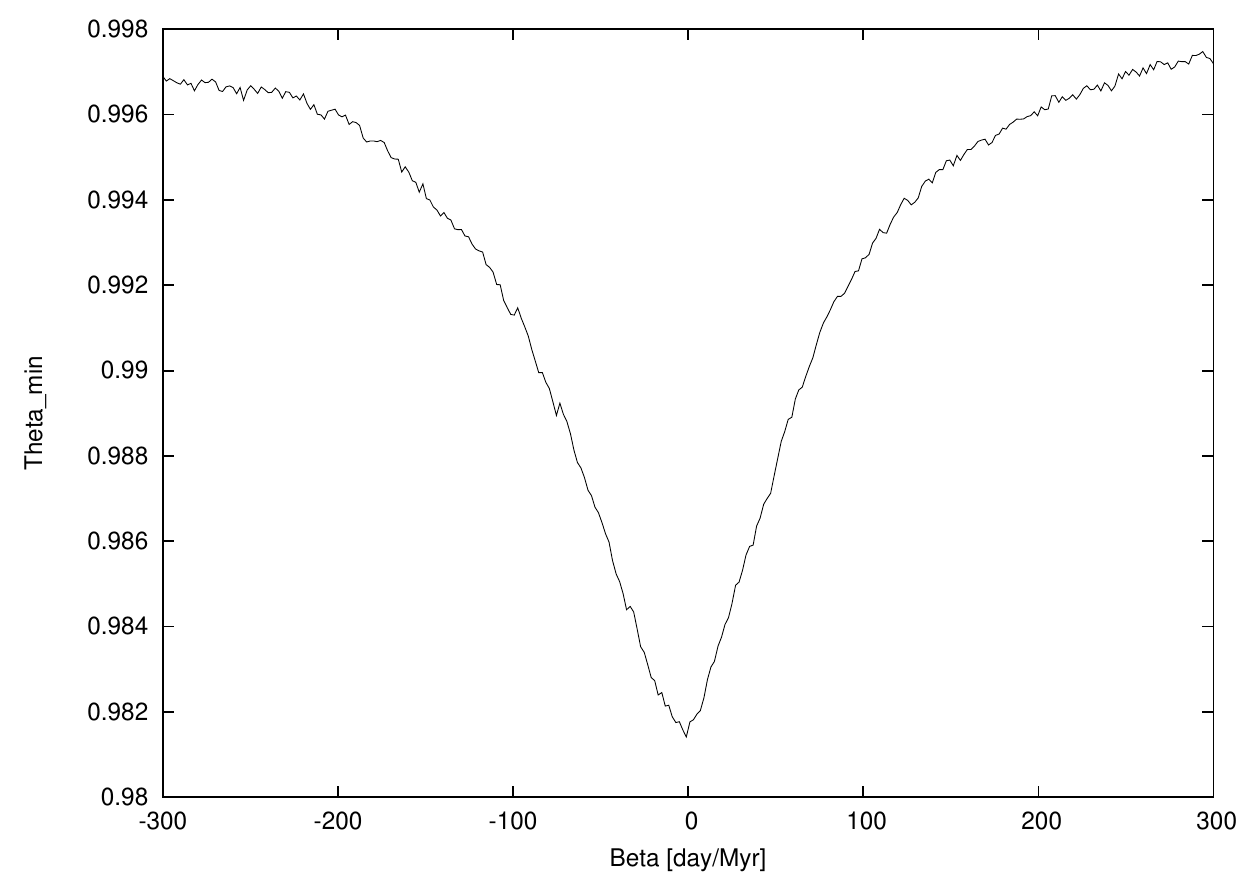}}
\caption{The light curve (left panel), transit residuals
(mid panel), and the long-term change of orbital period (right
panel) of the exoplanet candidate KIC8278371.}
\label{Fig. A15}
\end{figure*}

\begin{figure*}[!h]
\centering
\centerline{
\includegraphics[width=58mm]{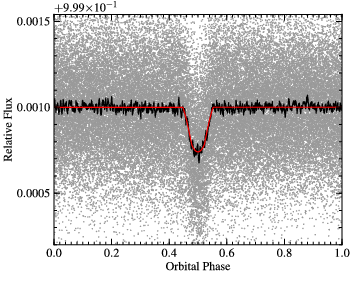}
\includegraphics[width=58mm]{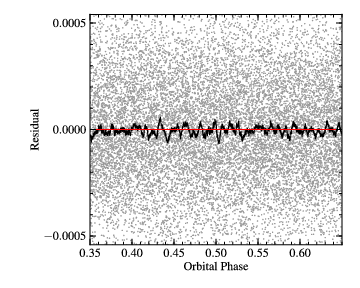}
\includegraphics[width=65mm]{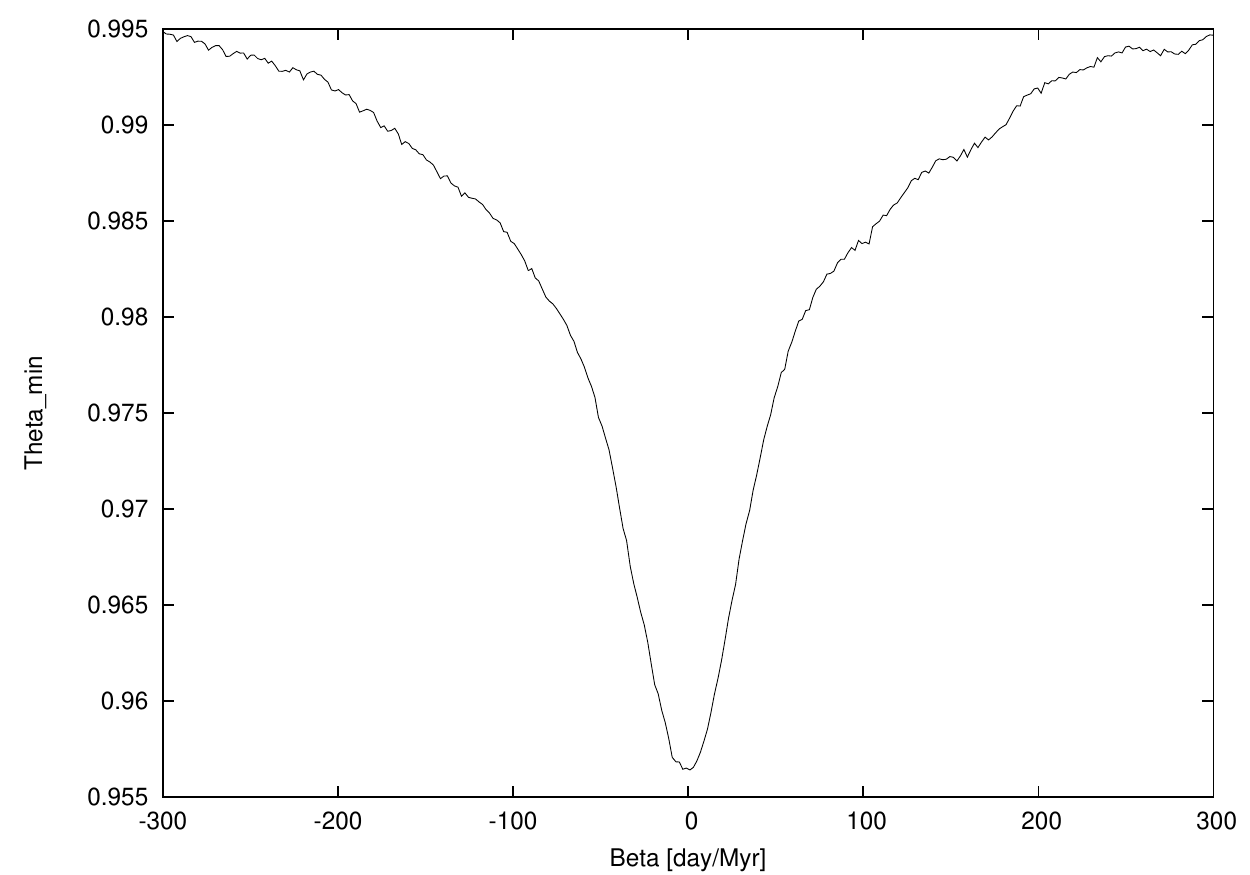}}
\caption{The light curve (left panel), transit residuals
(mid panel), and the long-term change of orbital period (right
panel) of the exoplanet candidate KIC5513012.}
\label{Fig. A16}
\end{figure*}

\begin{figure*}[!h]
\centering
\centerline{
\includegraphics[width=58mm]{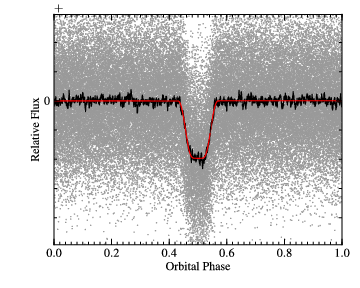}
\includegraphics[width=58mm]{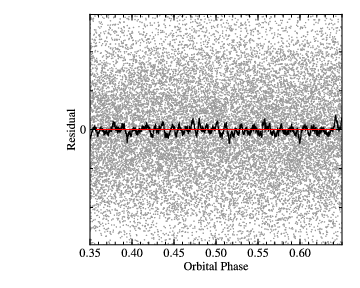} 
\includegraphics[width=65mm]{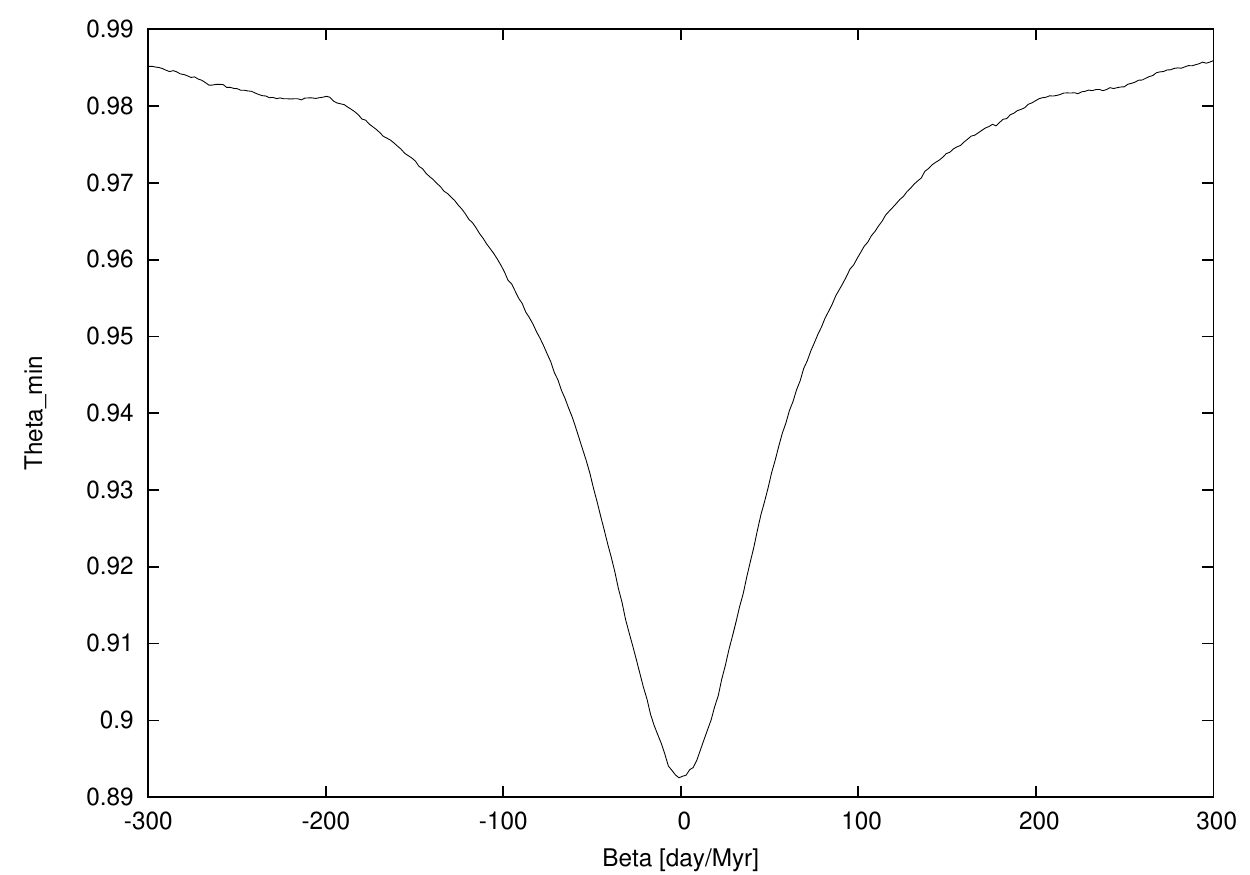}}
\caption{The light curve (left panel), transit residuals
(mid panel), and the long-term change of orbital period (right
panel) of the exoplanet candidate KIC10319385.}
\label{Fig. A17}
\end{figure*}

\begin{figure*}[!h]
\centering
\centerline{
\includegraphics[width=58mm]{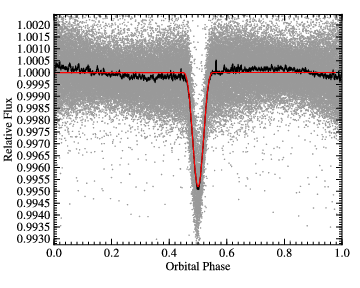}
\includegraphics[width=58mm]{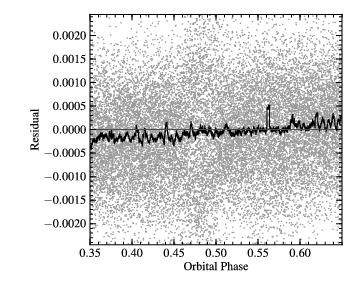}
\includegraphics[width=65mm]{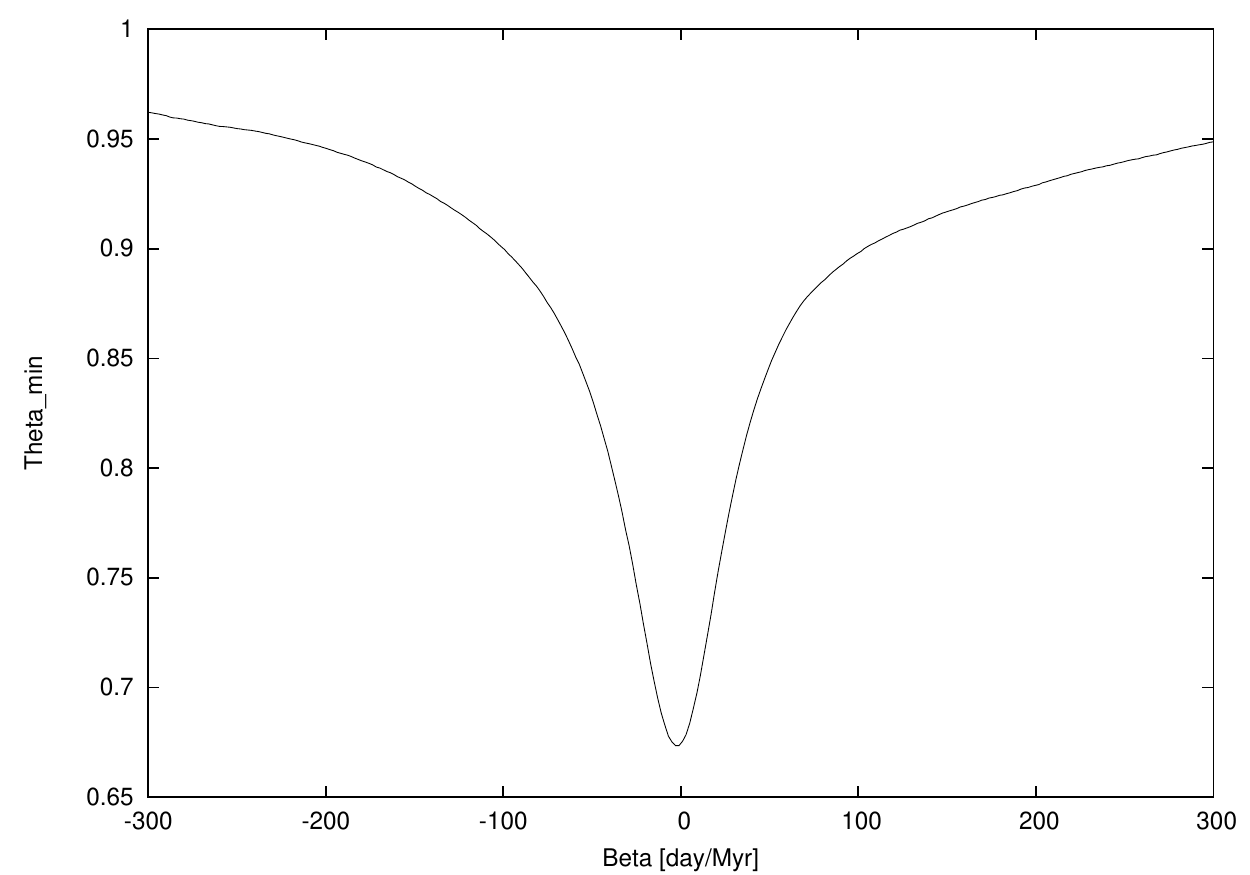}}
\caption{The light curve (left panel), transit residuals
(mid panel), and the long-term change of orbital period (right
panel) of the exoplanet candidate KIC9761199.}
\label{Fig. A18}
\end{figure*}

\begin{figure*}[!h]
\centering
\centerline{
\includegraphics[width=58mm]{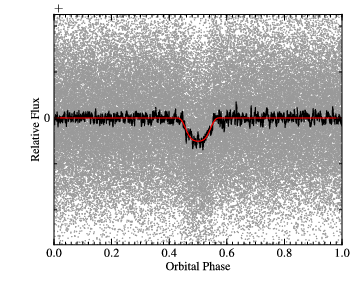}
\includegraphics[width=58mm]{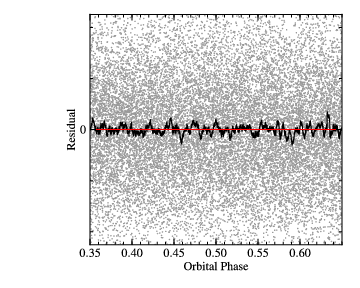}
\includegraphics[width=65mm]{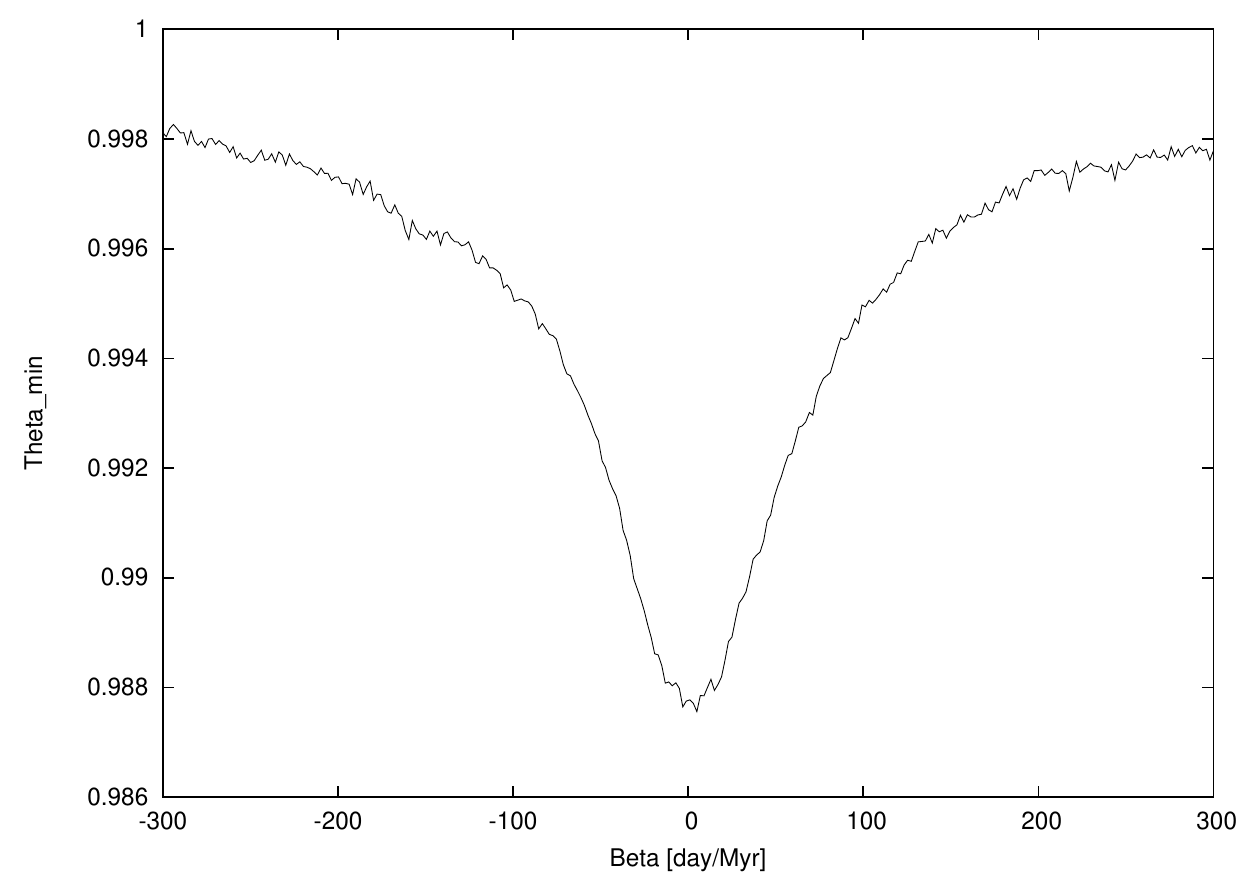}}
\caption{The light curve (left panel), transit residuals
(mid panel), and the long-term change of orbital period (right
panel) of the exoplanet candidate KIC9473078.}
\label{Fig. A19}
\end{figure*}

\begin{figure*}[!h]
\centering
\centerline{
\includegraphics[width=58mm]{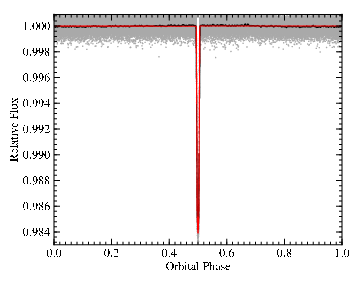}
\includegraphics[width=58mm]{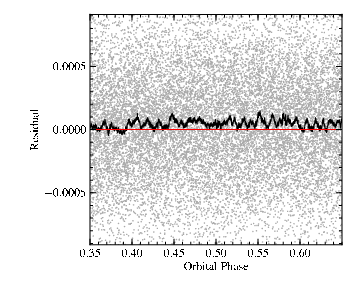}
\includegraphics[width=65mm]{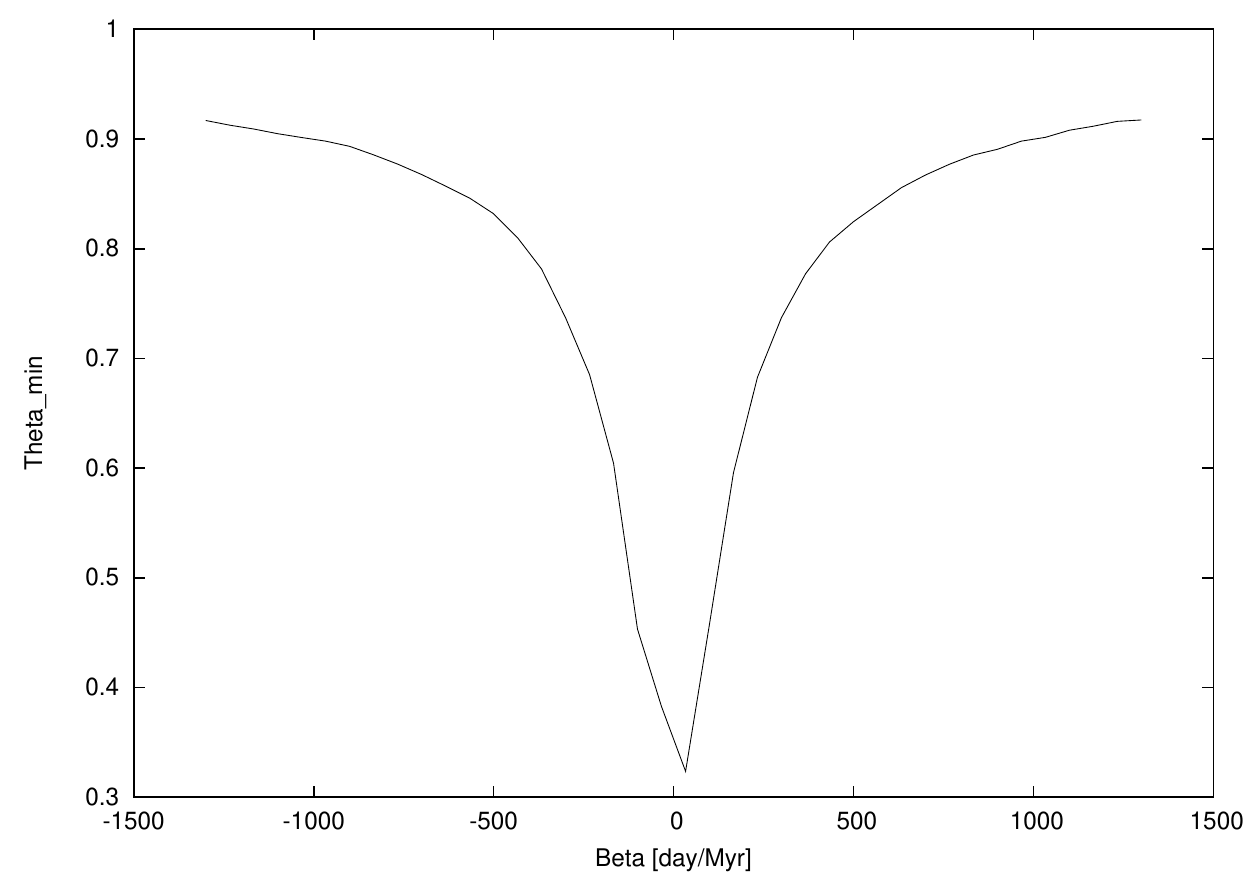}}
\caption{The light curve (left panel), transit residuals
(mid panel), and the long-term change of orbital period (right
panel) of the exoplanet candidate KIC5972334.}
\label{Fig. A20}
\end{figure*}


\newpage
\begin{thebibliography}{}

  \bibitem{} Agol, E., Steffen, J., Sari, R. \& Clarkson, W.: 2005, MNRAS 359, 567
  \bibitem{} Batalha, N.M., Rowe, J.F., Bryson, S.T., et al.: 2013, ApJS 204, 24
  \bibitem{} Bodenheimer, P., Lin, D.N.C. \& Mardling, R.A.: 2001, ApJ 548, 466
  \bibitem{} Borucki, W.J., Koch, D.G., Basri, G., et al.: 2011, ApJ 736, 19
  \bibitem{} Borucki, W.J., Koch, D.G., Batalha, N., et al.: 2012, ApJ 745, 120  
  \bibitem{} Bourrier, V., Lecavelier des Etangs, A., Dupuy, H., et al.: 2013, A\&A 551, 63
  \bibitem{} Brogi, M., Keller, C.U., de Juan Ovelar, M., Kenworthy, M.A., de
             Kok, R.J., Min, M. \& Snellen, I.A.G.: 2012, A\&A 545, 5
  \bibitem{} Budaj, J.: 2013, A\&A 557, 72
  \bibitem{} Burrows, A., Budaj, J. \& Hubeny, I.: 2008, ApJ 678, 1436
  \bibitem{} Burrows, A. \& Lunine, J.: 1995, Natur 378, 333
  \bibitem{} Col\'on, K.D., Ford, E.B. \& Morehead, R.C.: 2012, MNRAS 426, 342
  \bibitem{} Croll, B., Rappaport, S., DeVore, J., et al.: 2014, ApJ 786, 100
  \bibitem{} Esteves, L.J. De Mooij, E.J.W. \& Jayawardhana, R.: 2013, ApJ 772, 51
  \bibitem{} Faigler, S. \& Mazeh, T.: 2011, MNRAS 415, 3921
  \bibitem{} Ford, E.B., Fabrycky, D.C., Steffen, J.H., et al.: 2012, ApJ 750, 113
  \bibitem{} Ford, E.B., Rowe, J.F., Fabrycky, D.C., et al.: 2011, ApJ 197, 2
  \bibitem{} Foreman-Mackey, D., Hogg, D.W., Lang, D., \& Goodman, J.: 2013, PASP 125, 306 
  \bibitem{} Fortney, J.J., Lodders, K., Marley, M.S. \& Freedman, R.S.: 2008, ApJ 678, 1419
  \bibitem{} Fressin, F., Torres, G., Rowe, J.F., et al.: 2012, Natur 482, 195
  \bibitem{} Guillot, T., Burrows, A., Hubbard, W.B., Lunine, J.I. \& Saumon, D.: 1996, ApJ 459, 35
  \bibitem{} Holman, M.J. \& Murray, N.W.: 2005, Sci 307, 1288
  \bibitem{} Hubbard, W.B., Hattori, M.F., Burrows, A., Hubeny, I. \& Sudarsky, D.: 2007, Icar 187, 358
  \bibitem{} Hubeny, I., Burrows, A. \& Sudarsky, D.: 2003, ApJ 594, 1011
  \bibitem{} Jackson, B., Stark, C.C., Adams, E.R., Chambers, J. \& Deming,
             D.: 2014, AAS 223, 202
  \bibitem{} Kawahara, H., Hirano, T., Kurosaki, K., Ito, Y. \& Ikoma, M.: 2013, ApJ 776, L6
  \bibitem{} Kipping, D.M.: 2013, MNRAS 435, 2152 
  \bibitem{} Knutson, H.A., Charbonneau, D., Allen, L.E., Burrows, A. \& Megeath, S.T.: 2008, ApJ 673, 526
  \bibitem{} Kulow, J.R., France, K., Linsky, J. \& Loyd, R.O.P.: 2014, ApJ 786, 132
  \bibitem{} Kurokawa, H. \& Kaltenegger, L.: 2013, MNRAS 433, 3239
  \bibitem{} Lecavelier des Etangs, A., Ehrenreich, D., Vidal-Madjar, A. et al.: 2010, A\&A 514, 72
  \bibitem{} Loeb, A. \& Gaudi, B.S.: 2003, ApJ 588, 117
  \bibitem{} Lopez, E.D., Fortney, J.J. \& Miller, N.: 2012, ApJ 761, 59
  \bibitem{} Luddington, E.W.: 1978, BAAS 10, 418
  \bibitem{} Maciejewski, G., Dimitrov, D., Seeliger, M., et al.: 2013, A\&A 551, 108
  \bibitem{} Mandel, K. \& Agol, E.: 2002, ApJ 580, 171
  \bibitem{} Mazeh, T., Nachmani, G., Holczer, T., et al.: 2013, ApJS 208, 16
  \bibitem{} Ofir, A. \& Dreizler, S.: 2013, A\&A 555, 580
  \bibitem{} Owen, J.E. \& Wu, Y.: 2013, ApJ 775, 1050 
  \bibitem{} Rappaport, S., Barclay, T., DeVore, J., Rowe, J., Sanchis-Ojeda, R., Still, M.: 2014, ApJ 784, 40 
  \bibitem{} Rappaport, S., Levine, A., Chiang, E., et al.: 2012, ApJ 752, 1
  \bibitem{} Rybicki, G.B. \& Lightman, A.P.: 1979, Radiative Processes in
             Astrophysics, Wiley, New York 
  \bibitem{} Perez-Becker, D. \& Chiang, E.: 2013, MNRAS 433, 2294
  \bibitem{} Sanchis-Ojeda, R., Rappaport, S., Winn, J.N., Kotson, M.C., Levine, A.M., El Mellah, I.: 2014, ApJ 787, 47 
  \bibitem{} Slawson, R.W., Pr\'sa, A., Welsh, W.F., et al.: 2011, AJ 142, 160
  \bibitem{} Steffen, J.H., Batalha, N.M., Borucki, W.J., et al.: 2010, ApJ 725, 1226
  \bibitem{} Steffen, J.H., Ford, E.B., Rowe, J.F., et al.: 2012, ApJ 756, 186
  \bibitem{} Stellingwerf, R.F.: 1978, ApJ 224, 953
  \bibitem{} Stellingwerf, R.F.: 2004, PDM2, Stellingwerf Consulting, Huntsville 2004
  \bibitem{} Tassoul, J.L. \& Tassoul, M.: 1992, ApJ 395, 259
  \bibitem{} Tian, F., Toon, O.B., Pavlov, A.A. \& De Sterck, H.: 2005, ApJ 621, 1049
  \bibitem{} van Werkhoven, T.I.M., Brogi, M., Snellen, I.A.G. \& Keller, C.U.: 2014, A\&A 561, A3
  \bibitem{} Vidal-Madjar, A., D\'{e}sert, J.-M., Lecavelier des Etangs, A., et al.: 2004, ApJ 604, 69
  \bibitem{} Vidal-Madjar, A., Lecavelier des Etangs, A., D\'{e}sert, J.-M.,
             Ballester, G.E., Ferlet, R., H\'{e}brard, G. \& Mayor, M.: 2003, Natur 422, 143
  \bibitem{} Yelle, R.V.: 2004, Icar 170, 167
  \bibitem{} Zahn, J.P.: 1977, A\&A 57, 383

\end{thebibliography}
\end{document}